%% file: main.tex
\documentclass[acmsmall,10pt]{acmart}

\input{includes}

\begin{document}

\author{Thodoris Sotiropoulos}
\email{theosotr@aueb.gr}
\affiliation{%
  \institution{Athens University of Economics and Business}
}
\author{Stefanos Chaliasos}
\email{schaliasos@aueb.gr}
\affiliation{%
  \institution{Athens University of Economics and Business}
}
\author{Dimitris Mitropoulos}
\email{dimitro@aueb.gr}
\affiliation{%
  \institution{Athens University of Economics and Business}
}
\author{Diomidis Spinellis}
\email{dds@aueb.gr}
\affiliation{%
  \institution{Athens University of Economics and Business}
}

\settopmatter{printacmref=false}
\setcopyright{none}
\renewcommand\footnotetextcopyrightpermission[1]{}

\sloppy
\title{Identifying Bugs in Make and JVM-Oriented Builds}

\begin{abstract}

Incremental and parallel builds are
crucial features of modern build systems.
Parallelism enables fast builds by
running independent tasks simultaneously,
while incrementality saves time
and computing resources
by processing the build operations that were
affected by a particular code change.
Writing build definitions
that lead to error-free incremental
and parallel builds
is a challenging task.
This is mainly because developers
are often unable to predict
the effects of build operations
on the file system
and how different build operations
interact with each other.
Faulty build scripts may seriously degrade
the reliability of automated builds,
as they cause build failures,
and non-deterministic and incorrect build results.

To reason about arbitrary build executions,
we present~\bfs,
a generally-applicable model
that takes into account the specification
(as declared in build scripts)
and the actual behavior (low-level file system operation)
of build operations.
We then formally define different types of
faults related to incremental and parallel builds
in terms of the conditions
under which a file system operation violates
the specification of a build operation.
Our testing approach,
which relies on the proposed model,
analyzes the
execution of single full build,
translates it into~\bfs,
and uncovers faults by checking for
corresponding violations.

We evaluate the effectiveness,
efficiency,
and applicability of our approach
by examining 612 Make
and Gradle projects.
Notably,
thanks to our treatment of build executions,
our method is the first to handle Java-oriented
build systems.
The results indicate
that our approach is
(1) able to uncover several important issues
(245 issues found in 45 open-source projects
have been confirmed and fixed by the upstream developers),
and (2) up to six orders of magnitude faster
than a state-of-the-art tool for Make builds.

\end{abstract}

\maketitle

\section{Introduction}
\label{sec:intro}

Automated builds are an integral part
of software development.
Developers spend considerable amounts of time
on writing and maintaining
scripts~\cite{build-effort,build-maintenance}
that implement the build logic of their project.
Such scripts may involve
the compilation of source files,
application testing,
and the construction of software artifacts
such as libraries and executables.
The advent of Continuous Integration ({\sc ci})
together with the complexity of modern software systems
have made prominent two important
properties related to automated builds:
efficiency and reliability~\cite{decomposition,metamorphosis,building,ci}.
To save computing resources
and development time~\cite{ci,nandor2019},
build tools must be capable of coping
with complex systems quickly,
but without sacrificing the reliability
of the final deliverables.
Following this direction,
new build systems have emerged
providing features such as
parallelism~\cite{metamorphosis,bazel,apmake},
caching~\cite{gradle-cache},
incrementality~\cite{oopsla-incremental,visser},
and the lazy retrieval of project dependencies~\cite{molly}.

Among these features,
parallelism and incrementality
are in the heart
of almost every modern build system.
Parallel builds reduce build times
by processing independent build operations on multiple {\sc cpu} cores.
Incrementality saves time and resources
by executing only those build operations
affected by a specific change in the codebase.
Both features are vital for a smooth development process,
as they significantly shorten
feedback loops~\cite{visser,building}.
For example,
thanks to parallelism,
building huge systems,
such as the Linux Kernel or {\sc llvm},
which consist of million lines of code
and thousands of source files,
can complete in a few minutes.

Parallel and incremental builds
though,
pose threats to
the reliability of the
build process
when they are not used with caution.
Conceptually,
a build is a sequence of tasks
that work on some input files,
and produce results
(output files),
potentially used by other tasks.
To avoid failures and race conditions,
developers must specify all dependencies
in their build scripts,
so that the underlying build system
does not process dependent tasks in
the wrong sequence or in parallel
(e.g., linking before
compilation is erroneous).
Similarly,
for correct incremental builds,
developers need to enumerate all source files
that a build task relies on.
This ensures
that after an update to a source file,
all the necessary tasks are re-executed to
generate the new build artifacts
reflecting this change.
Build scripts are susceptible to faults
because declaring all task dependencies
is a challenging
and error-prone task~\cite{nandor2019,decomposition,google-debt}.
Even best practices~\cite{make-deps},
and tools~\cite{cmake}
for managing dependencies automatically
are often insufficient
for preserving correctness~\cite{nandor2019}.
Build failures,
non-deterministic and inconsistent build outputs,
or time-consuming builds,
are inevitably the result of
such faults~\cite{nandor2019,build-effort}.

There is little prior work focusing on
detecting incorrect build definitions,
and existing approaches suffer from two
major shortcomings~\cite{nandor2019,bee}
(as we discuss in~\ref{sec:motivation})
that prevent them from being useful
in practice.
First,
previous approaches
are tailored to analyze
Make-based builds~\cite{MAKE} only.
Therefore,
applying the technique behind existing tools
to other build systems,
such as Java-based build tools,
is not possible.
One of the main reasons is that prior work
makes strong assumptions
about the internal behavior
of build systems
that are only relevant to Make builds.
Unfortunately,
there are not any techniques available to examine
the reliability of builds
originated from other
systems beyond Make (e.g. Gradle),
even though such build systems
are extensively used~\cite{build-effort,java-builds}
and suffer from similar issues~\cite{gradle-bug,gradle-incremental}.
Second,
fault localization using prior methods
requires a large amount of time
that hinder their adoption.
For example,
employing {\it mkcheck}~\cite{nandor2019} to
detect build-related issues in
a Make-based project consisting of
a hundred files
can take hours (or even days).

We propose an
{\it effective}
and {\it efficient} dynamic method
for detecting faults in parallel
and incremental builds.
Our method is based on a model (\bfs)
that treats a build execution
stemming from an~\emph{arbitrary} build system
as a sequence of tasks,
where each task receives a set of input files,
performs a number of file system operations,
and finally produces a number of output files.
\bfs~takes into account (1) the specification
(as declared in build scripts)
and (2) the definition
(as observed during a build through file accesses)
of each build task.
By combining the two elements,
we formally define
three different types of faults
related to incremental and parallel builds
that arise when a file access violates
the specification of build.
Our testing approach operates as follows.
First,
it monitors the execution
of a build script,
and models this execution in~\bfs.
Our method then verifies
the correctness of the build execution
by ensuring that there is no file access
that leads to any fault concerning incrementality
or parallelism.
Note that to uncover faults,
our method only requires a single full build.

We demonstrate the applicability of our approach
on build scripts written in two popular build automation systems,
namely,
Make and Gradle.
Make is one of the most
well-established build tools~\cite{build-maintenance},
while Gradle is a modern Java-based system
that has become the de-facto build tool
for Android and Kotlin
programs~\cite{kotlin-book,gradle-android,android-study}.
Our approach is also applicable
to other build systems
such as Ninja, Bazel or Scala's sbt.
To the best of our knowledge,
our approach is the first
treatment of Java-oriented build executions.

\noindent
{\bf Contributions.}
Our work makes the following contributions.

\begin{itemize}
\item We propose~\bfs,
a model for specifying
and verifying arbitrary build executions,
which is the key for applying
our fault detection approach in
both traditional (e.g., Make),
and modern (e.g., Gradle) build systems
(Section~\ref{sec:modelling}).

\item We introduce a dynamic method
that relies on~\bfs,
and is able to
uncover issues in parallel
and incremental builds
by analyzing the execution of a single clean build.
(Section~\ref{sec:approach}).

\item We evaluate the effectiveness
and the applicability of our approach
by detecting issues in 324 out of 612
Make and Gradle projects.
Notably,
235 issues found in 45 open-source projects
were confirmed and fixed
by upstream developers.
Furthermore,
our approach is more effective,
and orders of magnitude faster
than the state-of-the-art
when analyzing Make projects
(Section~\ref{sec:evaluation}).

\end{itemize}

\noindent
{\bf Availability.}
We are planning to archive and make the source code
and data used in our experiments
publicly available.

\section{Background}
\label{sec:background}

We provide the basic elements of Make and Gradle.
Then,
we discuss the types of fault that may occur in corresponding
scripts and are related to incremental and
parallel builds.

\subsection{Build Systems}
\label{sec:build-systems}


{\bf Make.}
Make is the oldest build system
used today~\cite{MAKE,nandor2019}.
It provides a domain-specific language ({\sc dsl})
that allows developers
to write definitions of rules
that instruct the system
how to build certain targets.
For example,
the following rule states
that building the target {\tt source.o},
which depends on the file {\tt source.c},
requires to invoke the {\tt gcc} command
as shown at line 2.
\begin{lstlisting}[language=make]
source.o: source.c
    gcc -c $^
\end{lstlisting}

By default,
Make builds
every target incrementally,
meaning that it generates
targets only when
they are missing or
when their dependent files are more recent
than the target.
Make uses file timestamps
to determine whether a file has changed or not.
Also, it provides some built-in variables
starting with the symbol ``{\tt \$}''.
The most common ones are {\tt \$@}
and {\tt \$\textasciicircum},
which refer to the name of the target
(e.g., {\tt source.o}),
and the dependencies
(e.g., {\tt source.c})
of the current rule respectively.
Developers write their Make rules in
files called~\emph{Makefiles}.
In particular,
developers can either write their own Makefiles,
or for increasing their productivity,
they can use higher-level tools,
such as CMake~\cite{cmake}
or {\sc gnu} Autotools~\cite{autotools}
that automatically generate Makefile definitions.
CMake offers its own {\sc dsl},
and enables programmers to write rules
which in turn are translated into Makefiles.
CMake is useful for managing systems with complex structure.
Autotools is a collection of tools
that configure and generate Makefiles from templates.

{\bf Gradle.}
Although newer than other
Java-based build tools,
such as Ant and Maven,
Gradle has gained much popularity recently.
Currently,
around 55\%
of the most popular Java Github projects use Gradle~\cite{java-builds},
and it has become the preferred build tool
for Kotlin and Android
programs~\cite{kotlin-book,gradle-android,android-study}.
Gradle is at least two times faster
than Maven~\cite{gradle-maven},
as it offers features,
such as parallelism, and a build cache.

Gradle provides a Groovy-
and a Kotlin-based {\sc dsl}
which adopts a task-based programming model.
In this sense,
Gradle programmers assemble build logic
in a set of tasks.
A task is a fundamental component in Gradle
that describes a piece of work needed to be done
as part of a build.
Developers can impose constraints
on the execution order of tasks.
Then,
Gradle represents the build workflow
as a directed acyclic graph
and processes every task in topological ordering.
To enable incremental builds,
developers need to enumerate
the files consumed and produced by each task.
In this context,
a task is executed
only when there is a change to any
of its input or output files.
Gradle adopts a content-based approach
to identify updates:
it compares the checksum of the
input / output files
with that coming from the last build.
Consider the following snippet:

\begin{lstlisting}[language=gradle]
task extractZip {
   inputs.file "/file.zip"
   outputs.dir "/extractedZip"
   from zipTree("/file.zip")
   into "/extractedZip"}
\end{lstlisting}

\noindent
The listing above demonstrates
a task named {\tt extractZip}
written in Gradle.
This task extracts the contents of
an archive,
namely {\tt /file.zip},
into the directory {\tt /extractedZip}.
The input and the output files of this task
are declared at lines 2 and 3 respectively.
Declaring the input / output makes
the task {\tt extractZip} incremental.
In this context,
Gradle re-executes this task only when
any of those files are modified.
Notice that an input or an output file can be
a directory (See line 3).
In this case,
Gradle recursively examines the contents
of the directory for updates.

Gradle provides a rich {\sc api}
that developers can rely on to customize
their builds,
or create plugins.
A plugin consists of a set of common tasks
that can be reused across multiple projects,
e.g., consider a plugin that applies a linter to
the source files of a project.
Up to now,
there are more than~\nnum{3600} Gradle plugins
available for use~\cite{gradle-plugins}.

\subsection{Faults in Incremental \& Parallel Builds}
\label{sec:motiv-examples}

Three types of faults
can occur due to incorrect build definitions:
\emph{missing inputs},
\emph{missing outputs},
and~\emph{ordering violations}.
The first two are associated with incremental builds,
while the last one concerns parallelism.

\begin{figure}
\begin{lstlisting}[language=make]
CXXFLAGS=-MD
OBJS=CMetricsCalculator.o QualityMetrics.o
qmcalc: $(OBJS) qmcalc.o
    gcc -g qmcalc.o $(OBJS) -o $@
-include $(OBJS:.o=.d)
\end{lstlisting}
\vspace{-3mm}
\caption{
A Make definition
that does not capture the dependencies
of the object file {\tt qmcalc.o}.}
\label{fig:motiv-incr}
\vspace{-5mm}
\end{figure}

{\bf Missing Inputs.}
A build definition manifests a missing input issue,
when a developer fails to define
all input files of a particular build task.
This leads to faulty incremental builds,
because whenever there is an update to any of
the missing input files,
the dependent build task is not executed
by the build system.
Consequently,
the build system produces stale targets
and outputs.

As an example,
Figure~\ref{fig:motiv-incr}
shows the fragment of
a Make definition
taken from the {\tt cqmetrics} project.
This build creates the executable {\tt qmcalc}
by linking the object files
{\tt CMetricsCalculator.o},
{\tt QualityMetrics.o},
and {\tt qmcalc.o}
(lines 3--4).
Every object file is created
by a built-in Make rule
that compiles each implementation file {\tt .c}
with a command of the form
{\tt \$(CC) \$(CXXFLAGS) -c}.
By default,
the input file of these built-in rules
is only the underlying implementation file,
e.g., the input file of
the rule {\tt qmcalc.o} is {\tt qmcalc.c}.
However,
an object file might also depend on
a set of header files.
Thus,
changing a dependent header file
requires the re-generation of the object file.
The developers tackle this issue by compiling
every object file with the {\tt -MD} flag (line 1).
This flag stores all header files
that a target relies on
into a dedicated dependency file
whose suffix is {\tt .d}.
The developers include these dependency files
in their Makefile on line 5.
Although compiling source files with {\tt -MD}
follows the best practices for
managing Make dependencies automatically~\cite{make-deps},
the above script is faulty,
because only the dependency files
of the object files included
in the variable {\tt \$OBJS} (line 2) are considered.
The issue here is that
when there is an update to a header file
that the rule {\tt qmcalc.o} depends on,
the object file is not re-created.
Thus,
the final executable {\tt qmcalc} can be linked with
stale object files.
We reported this issue,
and the developers confirmed and fixed it.

{\bf Missing Outputs.}
A fault related to missing outputs
is similar to that related to missing inputs.
However,
this time
the cause of this problem is that
a developer does not properly enumerate
the output files of a task.
As with missing inputs,
this issue makes incremental builds
skip the execution of some build tasks
even if their outputs have changed.
Note also that Gradle caches
the output files of a task
from previous builds,
and reuses them in subsequent ones
when input files remain the same.
Hence,
missing outputs also affect
the performance of a build
making it run slower.

This is an important feature
that makes Gradle much more efficient
than other build systems~\cite{gradle-maven}.
Note that missing outputs do not appear in Make builds,
because Make considers only the timestamp
of input files to decide
if a target rule must be re-executed.

\begin{figure}
\begin{lstlisting}[language=gradle]
apply plugin: "java"
apply plugin: "com.github.johnrengelman.shadow"
shadowJar { classifier = "" }
\end{lstlisting}
\vspace{-3.5mm}
\caption{
A Gradle script with an ordering violation.}
\label{fig:motiv-paral}
\vspace{-6mm}
\end{figure}

{\bf Ordering Violations.}
Every build tool supporting parallelism
runs independent build tasks \emph{non-deterministically}.
This means
that the build system is free to process
unrelated operations in any order
for achieving high performance.
Non-determinism does not cause
any problems to the build process,
when two tasks are indeed independent,
and the one does not depend on the other.
However,
race conditions emerge,
when two build tasks are conflicting
(e.g, the one produces something
consumed by the other),
but are executed concurrently.
Developers can introduce ordering constraints
in their build definitions
as a side effect of explicitly defining dependencies
among conflicting build tasks.
An ordering violation occurs
when a developer does not specify
ordering constraints between two dependent tasks.
Note that an ordering violation does not relate
to the incremental issues discussed above.
That is,
there can be a task declared with
the correct input / output relations,
but it races
with another conflicting task.

Figure~\ref{fig:motiv-paral} shows
an example of an ordering violation.
Here we have an excerpt
of a real-world Gradle script
(from the {\tt nf-tower} project)
whose goal is to create the fat {\sc jar} of
a Java application.
A fat {\sc jar} packages
all {\tt .class} files of the current project
along with the {\tt .class} files of
project dependencies,
forming the executable distribution
of the project.
The code first applies
the built-in Gradle plugin {\tt "java"}
(line 1).
This plugin---among other things---runs
two tasks:
(1) the task {\tt classes}
that compiles all Java files into
their corresponding {\tt .class} files,
and (2) the task {\tt jar}
that generates a {\sc jar} file
containing only the classes of the current project.
In turn,
the code employs an external plugin
(line 2)
containing the task {\tt shadowJar}
(line 3)
that eventually generates
the fat {\sc jar} of the project.
The problem here is
that the name of the fat {\sc jar}
generated by the task {\tt shadowJar}
conflicts with the name of the naive {\sc jar}
produced by the task {\tt jar}.
The tasks {\tt jar} and {\tt shadowJar}
do not depend on each other,
so Gradle is free to schedule {\tt jar}
after {\tt shadowJar}.
This erroneous ordering
results in incorrect output,
i.e., the task {\tt jar} overrides
the contents of the {\sc jar} file
produced by the task {\tt shadowJar}.
A fix to this problem is
to create a fat {\sc jar}
with a different name
(e.g, changing its classifier at line 5 to {\tt "-all"}).
The developers of {\tt nf-tower} confirmed
and fixed this problem.

As we will see in Section~\ref{sec:evaluation},
such issues are widespread
and affect the reliability of
many software deliverables.
This motivates the design of
a generic approach for easing the
adoption of incremental and parallel builds
in practice.

\section{A Model for Build Executions}
\label{sec:modelling}

Designing a technique
that is able to locate faults in
incremental and parallel builds
regardless of the underlying build system
requires a generally-applicable
and precise model
for reasoning about build executions.
Existing models make assumptions about builds
that make the testing approaches
relying on them~\emph{ineffective}
when applied to certain build tools
(Section~\ref{sec:motivation}).
To address this,
we propose our model for understanding
build executions
(Section~\ref{sec:modelling2}).
Then,
we introduce the notion of task graph
(Section~\ref{sec:task-graph}),
a component that serves as a basis for
ensuring the correctness of a build execution
(Section~\ref{sec:correctness}).

\subsection{Motivation}
\label{sec:motivation}

Prior work on detecting faults
in Make incremental builds,
namely mkcheck~\cite{nandor2019},
models build execution as a set
of system processes created
by the build system
during execution,
e.g.,
Make creates
a new {\tt gcc} process
for compiling every source file.
This model treats every process
as a function that
takes an input,
and produces an output.
The input stands for
the set of files that are read
by the process,
while the output is the set of files
written by it.
The inputs and outputs
of every process are computed
by analyzing the system call trace of a build.
Through this model,
they infer the inter-dependencies among files
by considering each output to
be dependent on every input file.
All dependencies are then transitively
propagated using the process hierarchy.
However,
modeling build execution as a
set of system processes is problematic
for the following reasons.

{\bf Low Precision.}
The main assumption made by Make-based tools
is that the build system~\emph{always}
spawns a separate process
when proceeding to a new build task.
However,
this assumption is~\emph{no} longer valid
in modern build systems such as
Gradle, Maven, or Scala's sbt,
where the same system process
(e.g., JVM process)
involves multiple build tasks.
Tools that model builds as a sequence of processes
become ineffective
when applied to such build systems,
as their analysis precision significantly drops.

\begin{figure}[t]
\centering
\begin{subfigure}{.5\linewidth}
\centering
\includegraphics[width=.8\linewidth]{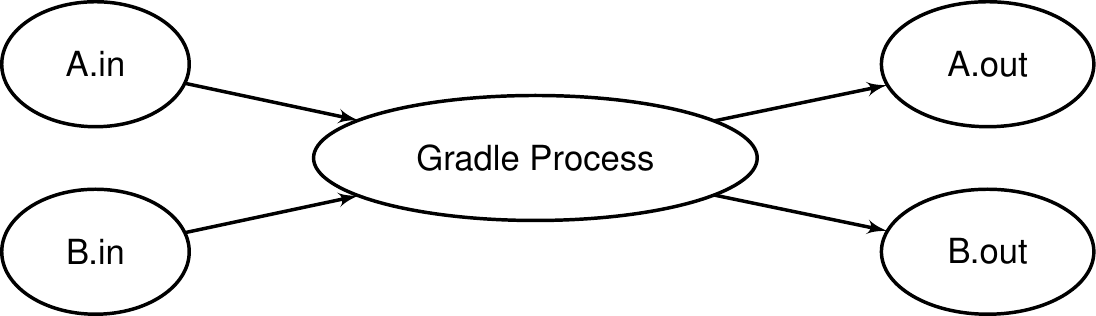}
\caption{Overconstrained graph}
\label{fig:overcon-graph}
\end{subfigure}%
\begin{subfigure}{.5\linewidth}
\centering
\includegraphics[width=0.7\linewidth]{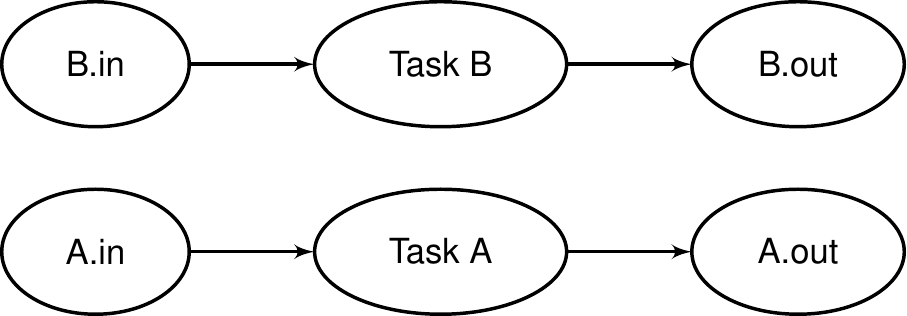}
\caption{Precise graph}
\label{fig:prec-graph}
\end{subfigure}
\vspace{-3mm}
\caption{Overconstrained dependency graph
computed by an approach modeling build execution
as sequence of processes vs.\
precise graph
produced by an approach modeling build execution
as a sequence of tasks.}
\label{fig:graphs}
\vspace{-5mm}
\end{figure}

To highlight how this feature of modern build systems
affect the precision of existing work we provide
a representative example.
Consider a Gradle task {\tt A} that reads a file
{\tt A.in} and creates a file {\tt A.out},
and a Gradle task {\tt B}
with file {\tt B.in} as its input,
and {\tt B.out} as its output.
An approach that works on granularity of processes
produces the dependency graph of Figure~\ref{fig:overcon-graph}.
The main Gradle process
runs both tasks A and B;
therefore, the analysis considers files
{\tt A.in} and {\tt B.in} as the inputs of that process,
and files {\tt A.out} and {\tt B.out} as its outputs.
Conceptually,
this~\emph{merges} the two tasks
into a single task.
The resulting graph is overconstrained~\cite{nandor2019},
because the analysis \emph{over-approximates}
the set of dependencies.
For example,
when there is a change in the input file {\tt A.in},
the analysis incorrectly considers that both
output files (i.e., {\tt A.out} and {\tt B.out})
must be updated,
even if {\tt A.in} only affects {\tt A.out}.
Overconstrained graphs lead to dozens of
false positives and negatives~\cite{nandor2019}.

{\bf Efficiency and Applicability.}
Another core limitation of the existing model
is that it only captures
OS-level facts (e.g., file accesses,
and file dependencies)
which are computed while analyzing the build trace.
To verify those inferred file dependencies
against the specification of build scripts,
prior work (i.e., mkcheck) triggers incremental builds
by touching each source file,
and checking whether the expected output files
are re-generated in response to the updated input files.
This makes the verification task extremely slow
as it requires substantial resources
when applying multiple incremental builds
in large-scale projects~\cite{nandor2019}
(see also Section~\ref{sec:eval-comparison}).

A critical reader may think
that combining static analysis
with dynamic analysis is a workaround for this
efficiency issue~\cite{bee}.
Specifically,
another approach could perform static analysis
on build scripts to extract task specification,
and then compare this specification against
the actual behavior of task observed
during build execution.
Nevertheless,
reliably extracting task specification
from build scripts through static analysis
is particularly challenging
(and in many cases not possible)
for multiple reasons~\cite{nandor2019}.
First,
static analysis cannot reason about tasks
whose inputs / outputs
are dynamically computed
and are not known in build scripts.
The same applies for tasks not explicitly
mentioned in build scripts,
e.g., tasks defined in external Gradle plugins
as illustrated in Section~\ref{sec:motiv-examples}.
Second,
static analysis needs to reason about
the complex semantics of build system's {\sc dsl}.
This limits generalizability,
as applying the approach to a new build system
requires implementing a new static analyzer
which involves a lot of engineering effort.
Third,
even when a static analyzer is available,
OS-level facts (inferred dynamically by the existing model)
are not comparable with task specifications
(computed statically),
when the build system abstracts tasks as
arbitrary functions as in the case of Gradle
or Scala's sbt.
To further clarify this,
consider the following example.

\noindent
\begin{minipage}{.45\linewidth}
\begin{lstlisting}[language=gradle]
task A {
  inputs.file "/file/A"
  // Arbitrary operation...
}
task B {
  inputs.files ("/file/A", "/file/B")
  // Arbitrary operation...
}
\end{lstlisting}
\end{minipage}
\hfill
\begin{minipage}{.45\linewidth}
\begin{lstlisting}[language=strace]
open("/file/A")
... // other file system operations
open("/file/B")
... // other file system operations
open("/file/A")
\end{lstlisting}
\end{minipage}

\noindent
In the Gradle script on the left,
we have two incremental tasks
(task {\tt A} and {\tt B})
performing some arbitrary operations.
The specification of the task {\tt A} says
that this task is expected to
consume the file {\tt /file/A},
while the task {\tt B} reads the files
{\tt /file/A} and {\tt /file/B}.
Note that
the specification only indicates the intent
of the developer,
and~\emph{not} the actual interactions
of task with the system.
The latter is shown in the execution trace on the right.
In this scenario,
it is not possible
to compare the actual behavior of tasks
(inferred by analyzing the execution trace on the right)
against build specification
(extracted statically from
the build script on the left).
This is because existing dynamic analysis techniques
are unable to map
the file accesses shown on the right to the task
they belong to.
This is necessary for deciding correctness.
For example,
if the first access comes from task {\tt A}
while the remaining ones stem from task {\tt B},
the build script is not faulty.
On the other hand,
if it is the other way around
(i.e., the last two accesses belong to the first task),
the task {\tt A} manifests a missing input
on file {\tt /file/B},
as it consumes a file
not mentioned in the build script.


\subsection{Modeling Builds}
\label{sec:modelling2}

\input{syntax}

All the points discussed in
the previous section are fundamental issues
associated with the method's design and underlying model,
and~\emph{not} with its implementation.
We introduce~\bfs,
a model for
thinking about build executions
that addresses the main limitations of existing work.

The proposed model treats every build
as a sequence of \emph{tasks}
rather than system processes.
Every task corresponds to the execution of
a build operation.
For example,
a task in \bfs~stands for the execution of
a target rule in Make and Ninja,
a goal in Java Maven,
or a Gradle task in Gradle.
This tackles low precision introduced by prior work,
because it enables us to relate
every build task to its correct input
and output files regardless of
the internal behavior of the build tool
(e.g., whether it spawns a separate process or not).
For example,
unlike the overconstrained dependency graph
of Figure~\ref{fig:overcon-graph},
\bfs~allows us
to infer the precise graph
shown in Figure~\ref{fig:prec-graph}.
\bfs~separates file accesses based on
which task they belong to.
Therefore,
it does not perform unnecessary merges
when encountering tasks governed by the
same process,
which is the main source of imprecision
in previous work~\cite{nandor2019}.

For dealing with efficiency and applicability,
\bfs~provides each task with
a specification that consists of
(1) a set of files that the task is expected to consume,
(2) a set of files that the task is expected to produce,
and (3) a set of task dependencies.
A task dependency indicates that
a task depends on another,
i.e., it is executed only
after the dependent task.
Beyond specification,
every task has a definition
containing all the (low-level)
file system operations
performed while executing the task,
e.g., reading and writing files,
or changing the OS transient structures,
such as the file descriptor table.
Combining the (high-level) specification
and actual behavior
(low-level file system operations) of each task
makes our approach efficient and applicable,
for we can verify correctness,
that is,
whether the actual behavior
conforms to the specification,
by analyzing a single clean build
i.e., no need to run incremental builds or
static analysis on build scripts.

Figure~\ref{fig:syntax} shows the complete
model for build executions.
A build execution
$b = \langle t^1, t^2\dots\rangle$
consists of a sequence of tasks.
Every task $t$ is described by a unique
name ($\tau \in$ \textit{TaskName}),
and contains a specification and a definition.
The specification declares the input / output files
and the dependencies of each task,
while its definition consists of statements.
For example,
\prim{task} {\tt A (/file/in): /file/out} \prim{after} $\bot = s$ means
that the task named {\tt A} consumes
the file {\tt /file/in},
produces the file {\tt /file/out},
has no dependencies (\prim{after} $\bot$),
while its definition is given by $s$.

A definition is one or more statements.
There are two types of statements.
First,
the~\prim{sysOp in} $z$ = $o$ statement executes
a system operation $o$ in a process given by $z$.
Every process defines a scope
for file descriptor variables (\prim{fd}$_f$).
File descriptor variables point to paths
and are used to model the
file descriptor table of Unix-like processes.
An operation ($o \in {\it Op}$) executed inside
a process $z$ may introduce
new file descriptor variables
in the current process (scope)
(\prim{let fd$_f =\dots$}),
or delete existing ones
(\prim{del}).
Moreover,
an operation may perform various file system updates,
including file creation (\prim{produce})
and file consumption (\prim{consume}).
An expression ($e \in Expr$) can be a constant path,
a file descriptor variable,
or $p$~\prim{at} $e$.
The latter allows us to interpret the path $p$
relative to the path given by the $e$ expression
(note that the result of an expression is a path).
Finally,
the~\prim{newproc} $z_1$ statement
creates a fresh process (scope) $z_1$,
and it optionally copies all file descriptor variables
of an existing process $z_2$ to $z_1$
(\prim{newproc} $z_1$~\prim{from} $z_2$).
This models process forking.

\begin{figure}
\begin{minipage}{.48\linewidth}
\begin{lstlisting}[language=make]
# Copying file using a Make rule.
target: "/source"
    cp $^ $@
\end{lstlisting}
\begin{lstlisting}[language=gradle]
// Copying file using a Gradle task.
task target {
  inputs.file "/source"
  outputs.file "/target"
  from file("/source")
  into file("/target") }
\end{lstlisting}
\vspace{-3mm}
\captionof{figure}{Copying the contents of {\tt source}
into {\tt target}.}
\label{fig:example-builds}
\end{minipage}
\hfill
\begin{minipage}{.48\linewidth}
\begin{lstlisting}[language=fstrace, mathescape=true]
task target ("/source"): "/target" after $\bot$ =
  newproc p
  sysOp in p =
    let fd$_3$ = "/source"
    consume(fd$_3$)
    let fd$_4$ = "/target"
    produce(fd$_4$)
    del(fd$_4$)
    del(fd$_3$)
\end{lstlisting}
\vspace{-3mm}
\captionof{figure}{Modeling the execution
of the task {\tt target}
that stems from Make and Gradle scripts
of Figure~\ref{fig:example-builds}.}
\label{fig:example-fstrace}
\end{minipage}
\vspace{-3mm}
\end{figure}

As an example of modeling,
consider a simple build scenario
where we want to copy the contents
of the file {\tt /source} into
the file {\tt /target}.
Figure~\ref{fig:example-builds} shows
how we can express
this using Make and Gradle.
When we execute these build scripts,
the build system first opens
the file {\tt /source},
reads its contents,
then opens the file {\tt /target},
and finally writes the contents of {\tt /source}
to the file descriptor corresponding
to the second file.
Figure~\ref{fig:example-fstrace} illustrates
how we model the execution
stemming from these scripts.
Every build consists of a single task
named {\tt target}.
This task consumes {\tt /source} to create {\tt /target}.
The definition of the task {\tt target} creates
a new process (line 2)
and uses it to execute all
the file-related operations performed
when running Make and Gradle (lines 3--9).
For instance,
the operation~\prim{let fd$_3 = $} {\tt /source}
creates a new file descriptor
(in the current process)
pointing to file {\tt /source},
while the operation at line 5 consumes
this file descriptor.
These operations model file opening.
On the other hand,
the operation~\prim{del(fd$_4$)}
deletes the given file descriptor
once the task closes the corresponding file
(line 9).

The semantics of~\bfs~are also
shown in Figure~\ref{fig:syntax}.
Every task $t \in \textit{Task}$ is evaluated
on a state $\sigma \in \textit{Proc} \rightarrow \textit{FileDesc} \rightarrow \textit{Path}$.
The state gives all file descriptor variables
defined in every process.
The result of a task evaluation
$\llbracket t \rrbracket_\sigma \in \sigma \rightarrow \sigma \times r$
is a new state,
and the set of files consumed and produced by this task,
i.e.,
$r \in \mathcal{P}(\textit{Path}) \times \mathcal{P}(\textit{Path})$.
Note that the projection
$r_{\downarrow_c}$ gives the set of files
consumed by the task,
while $r_{\downarrow_p}$ is the set of produced files.
Statements,
operations,
and expressions are evaluated accordingly.
Notably,
operations and expressions are evaluated
inside the scope (process)
where they take place,
i.e., $\pi \in \textit{FileDesc} \rightarrow \textit{Path}$.
As an example,
after evaluating the following~\bfs~task
on the state $\sigma = \bot$,
we get a new state
$\sigma' = z_1 \rightarrow (1 \rightarrow \texttt{"/f1"},
2 \rightarrow \texttt{"/f1/f3"})$,
while the set of consumed files $r_{\downarrow_c}$
is $\{\texttt{"/f1/f3"}\}$,
and the set of produced files $r_{\downarrow_p}$
is $\{\texttt{"/f2/f4"}\}$.

\vspace{-1.5mm}
\begin{lstlisting}[language=fstrace, mathescape=true]
task $\tau$ ("/f1"): "/f2" after $\bot$ =
  newproc $z_1$
  sysOp in $z_1$ =
    let fd$_1$ = "/f1"
    let fd$_2$ = "f3" at fd$_1$
    consume(fd$_2$)
    produce("f4" at "/f2")
\end{lstlisting}
\vspace{-2mm}

\subsection{Task Graph}
\label{sec:task-graph}

\begin{figure}[t]
\centering
\begin{subfigure}{.5\linewidth}
\centering
\begin{lstlisting}[language=fstrace, mathescape=true]
task $t_1$ ("/f1"): "/f2"
  after $\bot$ = $s_1$
task $t_2$ ("/f2"): $\bot$
  after $t_1$ = $s_2$
task $t_3$ ("/f3", "/f4"): ("/f2", "/f5")
  after $\bot$ = $s_3$
\end{lstlisting}
\label{fig:bexec}
\end{subfigure}%
\begin{subfigure}{.5\linewidth}
\centering
\includegraphics[width=0.9\linewidth]{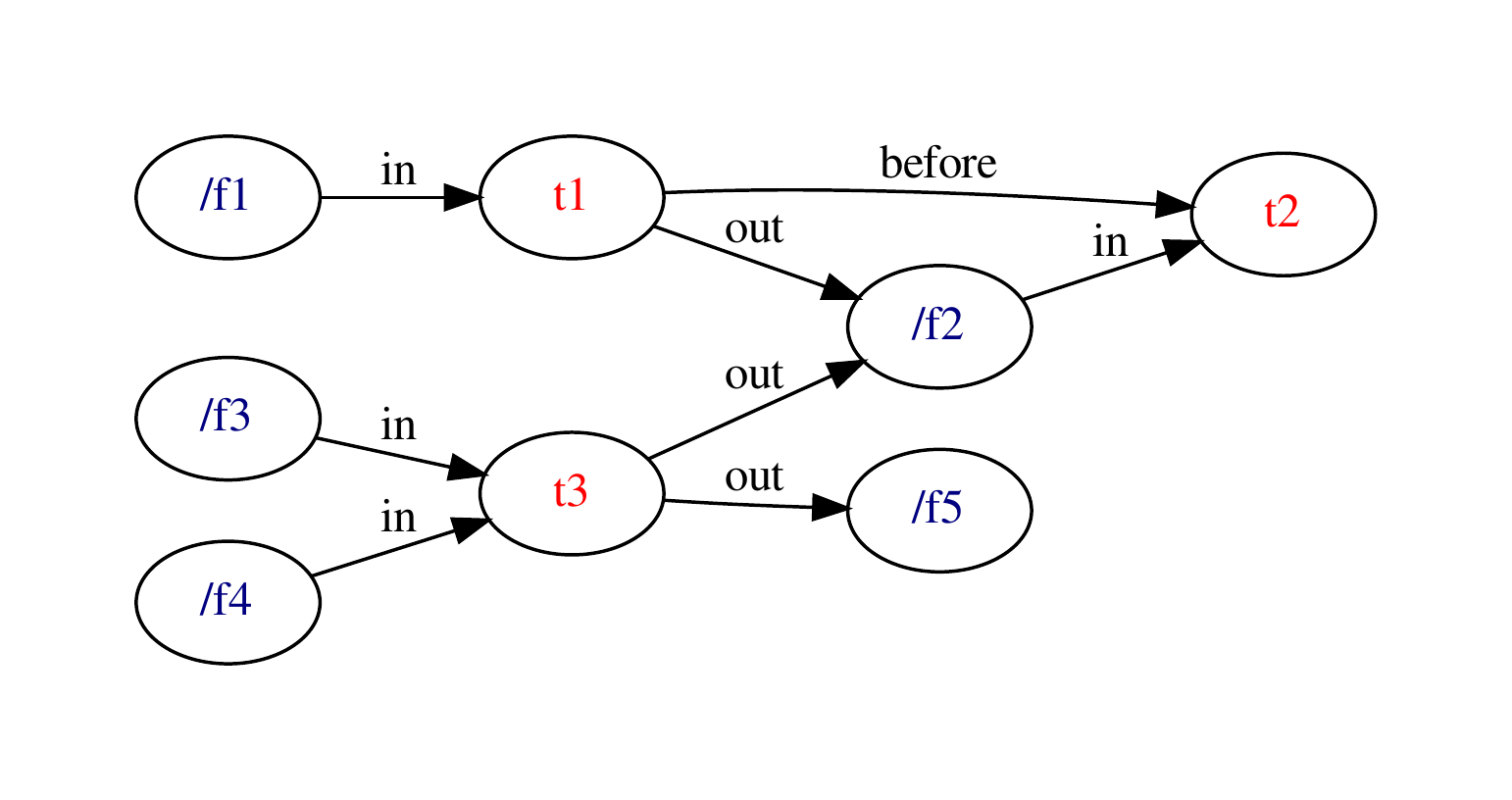}
\label{fig:tgraph}
\end{subfigure}
\vspace{-6mm}
\caption{An example of a build execution
and its task graph.}
\label{fig:task-graph}
\vspace{-5.5mm}
\end{figure}

Now,
we introduce the notion of~\emph{task graph}.
The task graph is a component
that stores the input files,
output files,
and the dependencies of every task
as declared by the developers in build scripts.
The task graph is computed by traversing
the specification of every task found
in a~\bfs~program (Section~\ref{sec:modelling2}),
and collecting all input/output files
and task dependencies.
We later use the task graph for
ensuring the correctness of a build execution
(Section~\ref{sec:correctness}).

We define the task graph as $G = (V, E)$.
A node $v \in V$ in the task graph is
either a task $t \in \textit{Task}$
or a file $p \in \textit{Path}$.
The set of edges $E \subseteq V \times V \times L$,
where $L = \{\prim{in}, \prim{out}, \prim{before}\}$,
determine the following relationships.
Given a task graph $G$,
the edge $p \overset{\primarr{in}}{\rightarrow}_G t$ indicates
that the file $p$ has been declared as an input
of the task $t$.
The edge $t \overset{\primarr{out}}{\rightarrow}_G p$
states that the task $t$ produces the file $p$.
Finally,
the edge $t_1 \overset{\primarr{before}}{\rightarrow}_G t_2$
shows a task dependency,
i.e.,
the execution of $t_1$ precedes that of $t_2$.

Given a build execution
modeled as a~\bfs~program
(Figure~\ref{fig:syntax}),
we gradually compute the task graph by
inspecting the specification of every task entry.
The $\overset{l}{\rightarrow}$ edges,
where $l \in \{\prim{in}, \prim{out}, \prim{before}\}$,
are constructed by
examining the header part of a~\prim{task} construct.
For instance,
whenever we encounter a task entry
of the form~\prim{task} $t_1$ ($p_1$): $p_2$~\prim{after} $t_2$,
we add the following edges to the task graph $G$:
(1) an $p_1 \overset{\primarr{in}}{\rightarrow}_G t_1$ edge,
(2) an $t_1 \overset{\primarr{out}}{\rightarrow}_G p_2$ edge,
and (3) an $t_2 \overset{\primarr{before}}{\rightarrow}_G t_1$ edge.

A complete example is shown in Figure~\ref{fig:task-graph},
where we have a build execution in~\bfs~on the left,
and its resulting task graph on the right.
Red nodes denote tasks while blue nodes indicate
files.

\subsection{Correctness of Build Executions}
\label{sec:correctness}

Having proposed our model for build executions
and the concept of task graph,
we now formalize the property of correctness
for build executions.
To do so,
we exploit the task graph
and define the~\emph{subsumption}
and~\emph{happens-before} relations
that we use as a base for
verifying correctness.


\begin{definition}
(Subsumption).
Given a task graph $G$,
we define the reflexive,
binary relation $\sqsubseteq_G$
and its transitive closure $\sqsubseteq_G^{+}$
on two paths $p_1, p_2 \in \textit{Path}$.
The definition is shown in Figure~\ref{fig:subsume}.
\end{definition}
\begin{figure}[t]
\centering
\small
\begin{mathpar}
\inferrule[self]{
    p \in \textit{Path}
}
{p \sqsubseteq_G p}
\hva \and
\inferrule[top]{
    p \in \textit{Path}
}
{p \sqsubseteq_G \top}
\hva \and
\inferrule[par-dir]{
    p_1, p_2 \in \textit{Path} \\
    \textsc{isParDir}(p_1, p_2)
}
{p_1 \sqsubseteq_G p_2}
\hva \and
\inferrule[indirect]{
    p_1, p_2 \in \textit{Path} \\
    t \in \textsc{nodes(g)} \\\\
    p_1 \sqsubseteq_G p_1'\\
    p_1' \overset{\primarr{in}}{\rightarrow}_G t \\
    t \overset{\primarr{out}}{\rightarrow}_G p_2
}
{p_1 \sqsubseteq_G p_2}
\hva \and
\inferrule[trans-clos]{
    p_1 \sqsubseteq_G p_2 \\
}
{p_1 \sqsubseteq_G^{+} p_2}
\hva \and
\inferrule[trans-clos]{
    p_1 \sqsubseteq_G^{+} p_2 \\
    p_2 \sqsubseteq_G^{+} p_3
}
{p_1 \sqsubseteq_G^{+} p_3}
\hva \and
\inferrule[mul]{
    t = (p_1, p_2\dots) \\
    \exists p' \in t\, p \sqsubseteq^{+}_G p'
}
{p_1 \sqsubseteq^{+}_G t}
\end{mathpar}
\vspace{-3mm}
\caption{Definition of the $\sqsubseteq_G$ relation through
inference rules.}\label{fig:subsume}
\vspace{-5mm}
\end{figure}

\begin{figure}[t]
\centering
\small
\begin{mathpar}
\inferrule[dep]{
    t_1, t_2 \in \textit{TaskName} \\
    t_1 \overset{\primarr{before}}{\rightarrow}_G t_2
}
{t_1 \prec_G t_2}
\hva \and
\inferrule[trans-clos]{
    t_1 \prec_G t_2 \\
}
{t_1 \prec_G^{+} t_2}
\hva \and
\inferrule[trans-clos]{
    t_1 \prec_G^{+} t_2 \\
    t_2 \prec_G^{+} t_3
}
{t_1 \prec_G^{+} t_3}
\end{mathpar}
\vspace{-3mm}
\caption{Definition of the $\prec_G$ relation through
inference rules.}\label{fig:happens-before}
\vspace{-3.5mm}
\end{figure}

The subsumption relation $p_1 \sqsubseteq_G p_2$ says
that the path $p_1$ is subsumed within the path $p_2$.
This relation is reflexive ({\sc [self]}),
and for every path $p \in \textit{Path}$,
we have $p \sqsubseteq_G \top$.
The relation $p_1 \sqsubseteq_G p_2$ holds
when $p_2$ is the parent directory of $p_1$
({\sc [par-dir]}),
or when $p_2$ relies on $p_1$,
i.e., there is at least one task in the task graph $G$
that produces $p_2$ using $p_1$
({\sc [indirect]}).
As we will see later
the subsumption relation is important for ensuring
that a file access made
while executing a build task
matches the task's specification.

\begin{definition}
(Happens-Before).
Given a task graph $G$,
we define the binary relation $\prec_G$
and its transitive closure $\prec_G^{+}$
on two tasks $t_1, t_2 \in \textit{Task}$.
The definition is shown in Figure~\ref{fig:happens-before}.
\end{definition}

The happens-before relation $t_1 \prec_G t_2$ states
that the task $t_1$ is executed before $t_2$.
The definition of this relation
consults the task graph $G$
to identify tasks
that are connected with each other
through an $\overset{\primarr{before}}{\rightarrow}_G$ edge,
which indicates a dependency between two tasks.
Finally,
the transitive closure
of $\prec_G$ gives
indirect task dependencies.
The happens-before relation enables us to verify
that two dependent tasks are
always executed in the correct order.

\subsubsection{Verifying Correctness of Tasks}
\label{sec:ver}

Using the subsumption relation,
we now formalize what the property of correctness
means for a~\bfs~task.

\begin{definition}
(Missing Input).
Given a task graph $G$
and a state $\sigma$,
a task $t \in \textit{Task} =$
\prim{task} $\tau$ $k_1$: $k_2$~\prim{after} $d = s$
manifests a~\emph{missing input} on state $\sigma$,
when
\begin{itemize}
    \item $(\sigma', r) = \llbracket t \rrbracket_{\sigma}$
    \item $\exists p \in r_{\downarrow_c}\, p \not\sqsubseteq_G^{+} k_1$
\end{itemize}
\end{definition}

\noindent
In other words,
to verify that a task
does not contain a missing input issue,
we first compute all file accesses
made by the task on
the given state $\sigma$,
i.e., $\llbracket t \rrbracket_{\sigma}$.
Then,
we check that every file consumed by this task
(i.e., $r_{\downarrow_c}$) matches
the input files $k_1$ declared
in the specification.
To do so,
we exploit the subsumption relation.
In particular,
when there exists a path $p$
consumed by this task
for which $p \not\sqsubseteq_G k_1$,
we say that the task has a missing input
on the state $\sigma$.
In practice,
this means that
although the build task relies on $p$
(as the definition of task consumes $p$),
the build system does not trigger
the execution of the task,
whenever $p$ is modified.

{\bf Example.}
Consider the following build execution
and its task graph.

\begin{minipage}{0.5\linewidth}
\centering
\begin{lstlisting}[language=fstrace, mathescape=true]
task $\tau_1$ ("/f1"): "/f2" after $\bot$ =
  newproc $z_1$
  sysOp in $z_1$ =
    consume "/f1/f3"
    produce "/f2"
task $\tau_2$ ("/f2"): $\bot$  after $\tau_1$ =
  sysOp in $z_1$ =
    consume "/f2"
    consume "/f1/f3"
    consume "/f3"
\end{lstlisting}
\end{minipage}
\hfill
\hspace{-5mm}
\begin{minipage}{0.5\linewidth}
\centering
\includegraphics[scale=0.45]{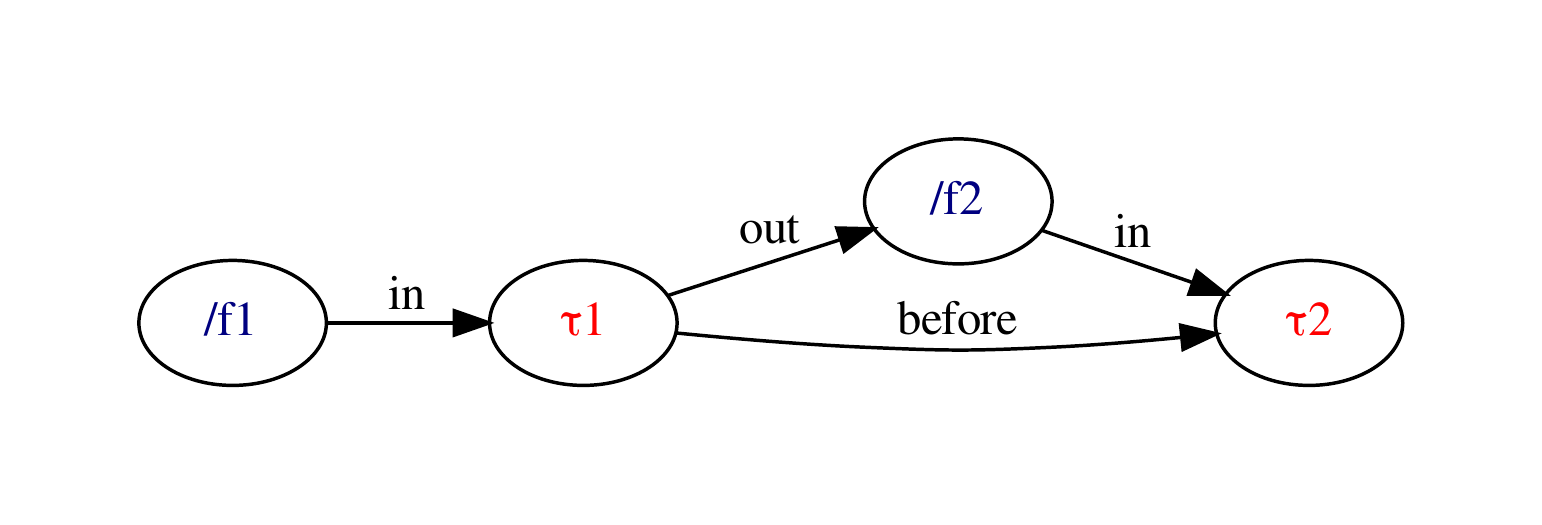}
\end{minipage}

When examining the task $\tau_1$,
we presume that it does not contain
any missing input issues,
as it only consumes the file {\tt "/f1/f3"}
(line 4)
which is subsumed within the input file {\tt /f1}
declared at line 1
(recall the {\sc [par-dir]} rule from Figure~\ref{fig:subsume}).
On the other hand,
the task $\tau_2$ consumes three files
(lines 8--10).
For the first access (line 8),
the $\sqsubseteq_G^{+}$ relation holds,
as the path consumed by $\tau_2$ is the same
with that declared in the specification.
The subsumption relation also holds for
the second access,
as the file {\tt f1/f3} is an input of the first task $\tau_1$
whose output is used
as an input for the target task $\tau_2$.
Therefore,
changing this file
will first trigger
the execution of task $\tau_1$.
This will eventually cause
the invocation of task $\tau_2$,
because the first task $\tau_1$ updates
the inputs of $\tau_2$.
This behavior is captured by the {\sc [indirect]}
rule (Figure~\ref{fig:subsume}).
Finally,
the task $\tau_2$ manifests a missing input
for the third access (line 10),
because we have ${\tt /f3}\not\sqsubseteq_G^{+}{\tt /f2}$.

\begin{definition}
\label{def:mout}
(Missing Output).
Given a task graph $G$
and a state $\sigma$,
a task $t \in \textit{Task} =$
\prim{task} $\tau$ $k_1$: $k_2$~\prim{after} $d = s$
manifests a~\emph{missing output} on state $\sigma$,
when
\begin{itemize}
    \item $(\sigma', r) = \llbracket t \rrbracket_{\sigma}$
    \item $\exists p \in r_{\downarrow_p}\, p \not\sqsubseteq_G^{+} k_2$
\end{itemize}
\end{definition}

The definition for missing outputs
is conceptually similar to that for missing inputs.
This time however,
we check that
for every file $p \in r_{\downarrow_p}$
produced by the examined task,
the $p \sqsubseteq_G^{+} k_2$ relation holds,
where $k_2$ stands for the declared output files
found in the specification of the task.

Given the above definitions,
we now introduce the notion of correctness for a
certain~\bfs~task.
\begin{definition}
(Correctness of Task).
Given a task graph $G$
and a state $\sigma$,
a task $t \in \textit{Task}$ is correct on state $\sigma$,
when it does not manifest a missing input
or missing output on state $\sigma$.
\end{definition}

\subsubsection{Verifying Correctness of Build Executions}
\label{sec:build-exec}

Recall
that a build execution in~\bfs~is
a sequence of tasks $b=\langle t^1, t^2\dots\rangle$.
A build execution may manifest
an ordering violation,
when there are pairs of tasks
that access a file $p$,
at least one of them produces $p$,
and there is no ordering constraint between
these tasks,
i.e., they can be executed in any order.

\begin{definition}
\label{def:ov}
(Ordering Violation.)
Given a task graph $G$
and an initial state $\sigma^0 = \bot$,
a build execution $b=\langle t^1, t^2\dots t^n\rangle$
manifests an~\emph{ordering violation} on
state $\sigma_i$,
when $\exists j \text{ with } 1\leq j<i < n$ such that
\begin{itemize}
    \item $(\sigma^{i+1}, r^{i+1}) = \llbracket t^{i+1} \rrbracket_{\sigma^{i}}$
    and $(\sigma^j, r^j) = \llbracket t^j \rrbracket_{\sigma^{j - 1}}$
    \item |$r^{i+1}_{\downarrow_p} \cap (r^j_{\downarrow_c} \cup r^j_{\downarrow_p})| \geq 1 \lor
        |r^j_{\downarrow_p} \cap (r^{i+1}_{\downarrow_c} \cup r^{i+1}_{\downarrow_p})| \geq 1$
    \item $t^{i+1} \not\prec_G^{+} t^j \land t^j \not\prec_G^{+} t^{i+1}$
\end{itemize}
\end{definition}

Contrary to missing inputs and outputs,
the definition for ordering violations
checks whether two tasks
with a conflicting file access
are executed in the right order.
To achieve this,
we use the happens-before relation $\prec_G^{+}$.
For example,
consider a task $t_1$
that creates a file $p$.
When the same file is consumed by a task $t_2$,
the $t_1 \prec_G^{+} t_2$ must hold.
Otherwise,
the build system is free to execute $t_2$ before
$t_1$.
Therefore,
$t_2$ may access a file
that does not exist,
resulting in a build failure.

\begin{definition}
\label{def:be}
(Correctness of Build Execution).
Given a task graph $G$
and an initial state $\sigma^0 = \bot$,
a build execution $b = \langle t^1, t^2\dots t^n\rangle$
is correct, when
\begin{itemize}
    \item the task $t^i$ is correct on state $\sigma^{i - 1}$,
        and when $(\sigma^i, r^i) = \llbracket t^i \rrbracket_{\sigma^{i - 1}}$
        the task $t^{i + 1}$ is also correct on state $\sigma^i$ for $1 \leq i < n$.
    \item the build execution does not manifest an ordering violation on
        state $\sigma^i$ for $1 \leq i < n$.
\end{itemize}
\end{definition}

Definition~\ref{def:be} summarizes our approach
for verifying a build execution.
We begin with examining and
evaluating tasks
in the order they appear in a~\bfs~program
according to the semantics of Figure~\ref{fig:syntax}.
The initial state is $\sigma_0 = \bot$.
Evaluating a build task gives us a new state,
and the set of files consumed
and produced by the task.
We then verify that the task is correct,
that is,
it does not contain any missing
inputs or outputs
while we also check
that it does not conflict with any previous task
based on the Definition~\ref{def:ov}
for ordering violations.
Finally,
we use the fresh state
to evaluate the next task
and perform the same verification task.

\section{Testing Approach}
\label{sec:approach}

\begin{figure}
\centering
\includegraphics[scale=0.50]{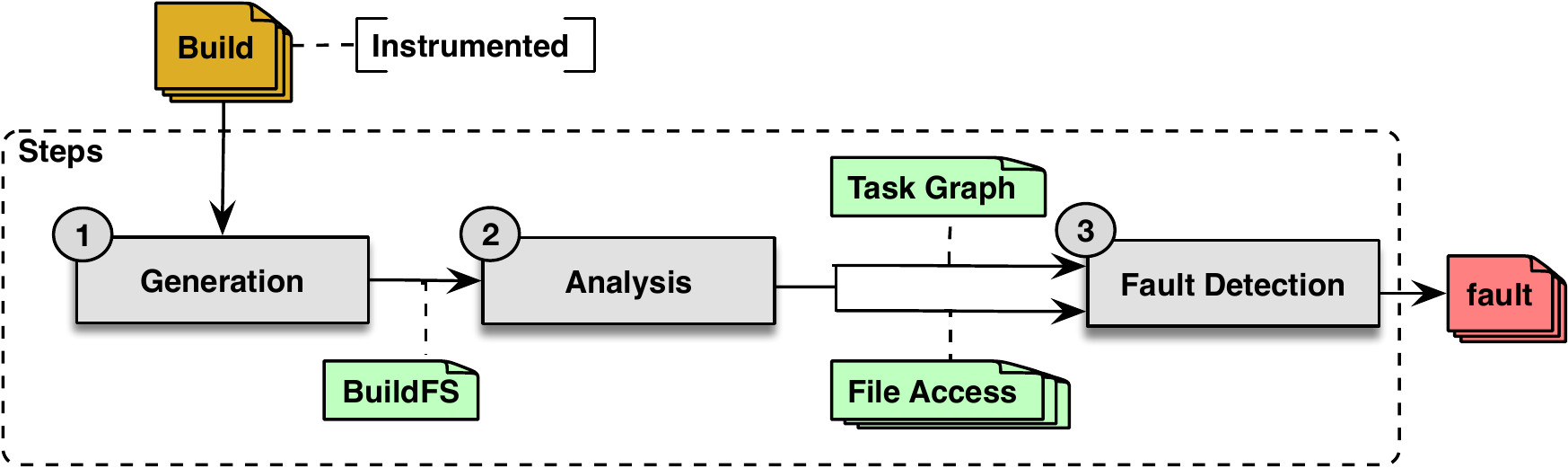}
\vspace{-2mm}
\caption{The steps of our testing approach.
It monitors the execution of an instrumented build,
and then it reports faults by analyzing
the generated~\bfs~programs.}
\label{fig:arch}
\vspace{-2mm}
\end{figure}

We now present the practical
realization of our model,
which works on the three phases
shown in Figure~\ref{fig:arch}.
During the first phase ({\it Generation})
we monitor the execution of an
instrumented build script,
and generate a~\bfs~representation
that models this execution.
As we will explain shortly,
the instrumentation performed on build scripts
provides the generation step with
all the necessary high-level information
to produce a valid~\bfs~program,
such as,
the execution boundaries
and specification of every build task.

In the second phase ({\it Analysis}),
we analyze the generated~\bfs~program,
and produce two outcomes.
First,
we construct the task graph
capturing the input / output files,
and the dependencies of every~\bfs~task.
Second,
we track all file accesses
of every task,
by evaluating each task $t^i$
in build execution
$b = \langle t^1, t^2\dots t^n\rangle$
as $\llbracket t^i \rrbracket_{\sigma_{i - 1}}$
for $1 \leq i \leq n$,
and with $\sigma^0 = \bot$.

The final step
({\it Fault Detection})
verifies the correctness of the given build execution
(modeled in~\bfs)
using the file accesses
and task graph
computed by the previous step.
Specifically,
this phase reports those file accesses
that violate the correctness of build execution,
i.e., they lead to missing inputs,
missing outputs,
or ordering violations
according to the definitions of Section~\ref{sec:correctness}.

\subsection{Generating BuildFS programs}
\label{sec:gen-traces}

\begin{figure}[t]
\begin{lstlisting}[language=strace]
+ write(1, "#BuildFS#: Begin target");
+ write(1, "#BuildFS#: target input /source");
+ write(1, "#BuildFS#: target output /target");
open("/source", O_RDONLY) = 3;
open("/target", O_WRONLY|O_CREAT) = 4;
read(3, "content");
write(4, "content");
close(4);
close(3);
+ write(1, "#BuildFS#: End target");
\end{lstlisting}
\vspace{-3mm}
\caption{Example of the native function calls observed,
while executing the build scripts of Figure~\ref{fig:example-builds}.
The highlighted function calls are those inserted by
our instrumentation.}
\label{fig:traces}
\vspace{-5mm}
\end{figure}

To model a build execution in~\bfs,
our dynamic approach
takes an~\emph{instrumented} build script as input
and monitors its execution.
The goal of the instrumentation
applied to build scripts is to
provide the execution boundaries of every task,
and other information coming from build definitions
(i.e., declared input / output files and dependencies).
To do so,
we place instrumentation points
before and after the execution of each build task,
and augment their execution
by calling special native functions.
These functions take a string argument
that either contains the information
originated from build definitions,
or indicate when the execution of a task begins or ends.
Then,
our dynamic analysis identifies these calls,
and extracts their arguments to construct
the specification of each task,
and map the intermediate file operations
to the corresponding task.
In this manner,
through monitoring these special native functions calls
and all the other file system operations
(e.g., a call to {\tt open}) that take place
while building,
we are able to construct~\bfs~programs.

An example of native function calls
inserted by our instrumentation are
writes to standard output.
These calls are triggered by inserting
simple {\tt print} statements
as part of the instrumentation.
Consider again the build scripts of
Figure~\ref{fig:example-builds}.
When monitoring their naive execution
(no instrumentation is added),
we observe the file system operations at lines 4--9
that reflect file copying.
When instrumenting these scripts,
we augment their execution
by adding the native function calls at lines 1--3,
before the execution of the task {\tt target},
along with the function call at line 10
after file copying.
These calls enables us to identify
when the task {\tt target} begins
and ends (lines 1, 10),
along with its input / output files (lines 2, 3).
Our dynamic analysis detects
and examines these calls,
and finally produces the~\bfs~representation
shown in Figure~\ref{fig:example-fstrace}.
Note that without this instrumentation,
we are unable to map the file system operations
(lines 4--9) to the task they come from,
and to build the specification of the
currently executed~\bfs~task.

The instrumentation
extracts task specifications
from the execution engine
of the build system,
and~\emph{not} from build scripts.
To perform a sound execution,
every build system is aware of all
the dependencies and input / output files
of each task at runtime.
For example,
the build system can recognize all task dependencies
to schedule the execution of a task
in the correct order.
Similarly,
in an incremental build,
the build system is aware of all declared file inputs
to determine which tasks must
be executed in response to some identified
file updates.
Our approach benefits from extracting this information
from the execution engine of build systems at runtime
for two reasons.
First,
we do not have to perform static analysis
(a challenging task as we discussed
in~\ref{sec:motivation})
to extract this information from build scripts.
Second,
we can recognize all dependencies and input / output
files,
including the ones computed dynamically
and not explicitly mentioned in build scripts.
Details regarding the instrumentation
of build scripts
are implementation specific
and depend on the underlying build system
as we show in Section~\ref{sec:implementation}.

\subsubsection{From File System Operations to BuildFS Operations}
\label{fs}
Here we describe
how we model a file-system operation
to a~\bfs~statement, operation or expression.

{\bf Operations on paths.}
A system operation that works on paths
is translated to either~\prim{consume}
or~\prim{produce} operations,
depending on its effect on the file system.
For example,
when a build task creates a new directory
through the {\tt mkdir("/dir")} system call,
we emit a~\prim{produce("/dir")} operation.
Similarly,
when the build system creates a hard link
to an existing file
by invoking the {\tt link("/source",} {\tt "/target")}
system call,
we yield two~\bfs~operations:
\prim{consume("/source")},
\prim{produce("/target")}.

{\bf Operations on file descriptors.}
When the build process creates a new file descriptor,
we use the~\prim{let fd$_f =\dots$} operation.
For example,
when creating a new file descriptor through opening a file
{\tt open("/file") = 3},
we emit~\prim{let fd$_3 =$ "/file"}.
We do the same,
when copying an existing file descriptor to a new one,
e.g., {\tt dup2(3, 4)} turns into
\prim{let fd$_4 =$ fd$_3$}.
Finally,
closing a file descriptor leads to~\prim{del} operations.

{\bf Working Directory.}
Each system process operates on a specific directory.
In~\bfs,
we use a special file descriptor
variable,
namely
\prim{fd$_0$},
that points the working directory of the current process.
Whenever,
the working directory of a process changes
(through the {\tt chdir} system call),
we emit a~\prim{let fd$_0 =\dots$} operation
to model this effect.

{\bf Relative Paths.}
Some file system operations operate on relative paths.
For example,
the call to {\tt mkdir("dir")}
creates the directory {\tt dir} inside
the current working directory.
We handle relative paths through
the~\prim{$p$ at $e$} expression.
Specifically,
we model the above example
as~\prim{produce("dir" at fd$_0$)}.

{\bf Forking Processes.}
When the build system creates
a new process from an existing one
(e.g., by calling the {\tt clone} system call),
we generate a~\prim{newproc $z_2$ from $z_1$} statement,
where $z_1$ refers to the {\sc id}
of the new process,
while $z_2$ is the parent process.
Finally,
we model the main build process
using a~\prim{newproc $z$} statement.

\subsection{Analyzing BuildFS Programs \& Detecting Faults}
\label{sec:analysis}

After modeling a build execution
in a~\bfs~representation,
our method performs a linear pass
over the representation and
produces two types of output.
First,
it generates the corresponding task graph,
and second,
it computes all file accesses
that take place in every~\bfs~task
based on the semantics presented
in Figure~\ref{fig:syntax}.

In the final step of our method
(fault detection),
we verify the correctness of
a build execution
based on the task graph and the file accesses
computed in the analysis step.
In particular,
we examine the file accesses of every task $t$,
and we proceed as follows.
If a file access $p$ is of type ``consumed'' (``produced''),
and the subsumption relation
(Section~\ref{sec:ver})
between $p$ and the file inputs (outputs) of $t$
does not hold,
we report a missing input (output) on $p$.
For ordering violations,
we check whether $p$ was accessed elsewhere
(say another task $t'$)
in the given~\bfs~program.
If this is the case,
we verify whether the execution order
between $t$ and $t'$ is deterministic
using the happens-before relation.
If the happens-before relation
between these tasks is undefined,
we report an ordering violation.
Our fault detection approach
eventually reports all file accesses
that violate the correctness of
the given build execution
according to the definitions of~\ref{sec:correctness}.

The faults related to parallelism manifest themselves
non-deterministically
(depending on the execution schedule of the build system),
while the ones associated with incrementality
do not appear in full builds.
However,
our technique is capable of detecting subtle
and future latent faults,
because it does not require the build
to crash and then reason about
the root cause of the failure.

\subsection{Implementation}
\label{sec:implementation}

\begin{figure}[t]
\begin{subfigure}{.6\linewidth}
\begin{lstlisting}[language=kotlin]
// Runs BEFORE the execution of every task.
fun processTaskBegin(task: Task) {
  task.inputs.files.forEach { input ->
  println("#BuildFS#: ${task.name} input ${input}")
 }
 task.outputs.files.forEach { output ->
  println("#BuildFS#: ${task.name} output ${output")
 }
 task.getTaskDependencies().forEach { d ->
  println("#BuildFS#: ${task.name} after ${d.name}")
 }
 println("#BuildFS#: Begin ${task.name}")
}
// Runs AFTER the execution of every task.
fun processTaskEnd(task: Task) {
 println("#BuildFS#: End ${task.name}")
}
\end{lstlisting}
\vspace{-2.5mm}
\caption{Fragment of the instrumentation applied to Gradle builds.}
\label{fig:kotlin}
\end{subfigure}
\begin{subfigure}{.39\linewidth}
\begin{lstlisting}[language=strace]
target=$1
prereqs=$2
taskName=$(pwd):$target
echo "#BuildFS#: Begin $taskName"
echo "#BuildFS#: $taskName input $prereqs"
shift 2
/bin/bash "$@"
echo "#BuildFS#: End $taskName"
\end{lstlisting}
\caption{The {\tt fsmake-shell} that instruments every {\tt Make} rule.}
\label{fig:make}
\end{subfigure}
\vspace{-3mm}
\caption{The instrumentation implemented for Gradle and Make builds.}
\vspace{-5mm}
\end{figure}

We have implemented our method
as a command-line OCaml program,
which we plan to make publicly available
as an open-source software.
To trace system operations,
we employ {\tt strace}~\cite{MMG06}.
Note that
this can be accomplished by using
either other system call tracing utilities
(e.g., {\tt DTrace}~\cite{Rod86}))
or dynamic binary
instrumentation~\cite{dynamo,valgrind}.

Our tool parses the {\tt strace} output
and translates it into a~\bfs~representation.
The implementation supports two modes:
\emph{offline},
and \emph{online}.
In the offline mode,
our tool does not monitor builds.
Instead,
it expects a file containing
the {\tt strace} output obtained from previous runs.
When in online mode,
the tool generates and analyzes a~\bfs~program,
while monitoring a build command
through {\tt strace}.
To do so,
it creates two processes.
The first process runs
{\tt strace} on the build command,
while the second reads the {\tt strace}
output produced by the first process
and runs the~\bfs~generation
and analysis steps
in a streaming fashion.
Communication is done through pipes which
allows processes to run concurrently.
Notably,
this eliminates the observable time
spent on the analysis phase,
because running the build is much slower
than the analysis of the corresponding
\bfs~programs.
Therefore,
in a multicore architecture,
our tool exploits a spare core to perform the analysis
as the build runs.

To instrument Gradle scripts,
we have implemented a Gradle plugin
written in Kotlin
that hooks~\emph{before} and~\emph{after}
the execution of every task
as shown in Figure~\ref{fig:kotlin}
(lines 1, 14 -- irrelevant code is omitted).
The plugin utilizes the Gradle
{\sc api}~\cite{gradle-custom-plugins}
to print the following elements:
(1) declared inputs / outputs of every task
(lines 3--5, 6--8),
(2) declared dependencies of every task (lines 9--11),
and (3) execution boundaries of every task (lines 12, 16).
This output is identified
by our dynamic analysis and
converted to~\bfs~tasks as explained in~\ref{sec:gen-traces}.
To apply our plugin to a Gradle project,
we modify Gradle scripts
by inserting~\emph{only} four lines of code.

For instrumenting Make scripts,
we created a shell script
({\tt fsmake-shell})
that~\emph{wraps} the execution of every Make rule
(Figure~\ref{fig:make}).
As with Gradle,
this script prints
the execution boundaries (lines 4, 8)
and prerequisites of each task (lines 5).
To achieve this,
we override Make's built-in variable {\tt \$SHELL}
to point to our script.
After printing the necessary information,
our script invokes the underlying shell
to eventually execute the requested Make command (line 7).
To handle Make dependencies
generated at build time
(e.g., through {\tt gcc -MD}),
we refine the task graph computed
during the analysis phase
by adding missing edges.
To do so,
we exploit information stemming from the Make database
by running {\tt make -pn} after each build.
Note that we do not need to make any changes
in the source code of build scripts to enable tracing;
we simply build projects by running
\begin{lstlisting}[language=make,numbers=none]
make "$@" -- SHELL='fsmake-shell '\''$@'\'' '\''$+'\'''  '''
\end{lstlisting}

Applying our method to a new build tool
requires little development effort.
Our Gradle plugin
contains 90 lines of Kotlin code,
while {\tt fsmake-shell} consists of
\emph{only 8 lines of shell code}.

{\bf Limitations.}
Currently,
our tool can trace builds
only in a Linux environment.
However,
extending our implementation
to support monitoring in other
platforms is straightforward.
Also,
{\tt strace} introduces a 2x times
slowdown on builds,
on average
(see Section~\ref{sec:eval-performance}).
Employing a tracing utility
that runs in the kernel space
to track system operations,
may reduce the overhead on build execution~\cite{jvm}.
Furthermore,
non-deterministic builds
(i.e., touching different files on different days)
may lead to false negatives
when a faulty file access does not happen
when running the build.
However,
unlike other approaches (e.g., mkcheck),
our tool can cope with non-determinism
occurred in subsequent builds
(e.g., temporary files
generated with random names),
because it requires a single
build for performing the verification.

\section{Evaluation}
\label{sec:evaluation}

We evaluate our approach
by answering the following
research questions:

\begin{enumerate}[label={\bf RQ\arabic*}, leftmargin=2.1\parindent]
\item {\bf (Effectiveness)} What is the effectiveness of our approach
in locating faults in build scripts?
(Section \ref{sec:eval-faults})
\item {\bf (Fault Importance)} What is the perception of developers
regarding the detected faults?
(Section~\ref{sec:eval-devs})
\item {\bf (Fault Patterns)} What are the main fault patterns?
(Section \ref{sec:eval-patterns})
\item {\bf (Performance)} What is the performance of our approach?
(Section \ref{sec:eval-performance})
\item {\bf (Comparison with state-of-the-art)}
How does the proposed approach perform with regards
to other tools (i.e., mkcheck)?
(Section~\ref{sec:eval-comparison})

\end{enumerate}
\subsection{Experimental Setup}
\label{sec:setup}

We applied our approach to
a large set of Gradle and Make projects.
To identify interesting Gradle projects,
we employed the
Github {\sc api}
to search for popular Java, Kotlin,
and Groovy repositories
that use Gradle.
We selected 200 projects
for each language
(i.e., 600 projects in total)
ordered by the number of stars.
For every project,
we performed the following steps.
First,
we instrumented the Gradle scripts
as described in Section~\ref{sec:implementation}.
Then,
we ran the instrumented Gradle scripts
through the {\tt gradle build} command,
which is the de-facto command for building Gradle projects.
Note that this command executes
the compilation, assembling and testing tasks
as well as other user-defined tasks.
For efficiency, we ran our tool in online mode
(Section~\ref{sec:implementation}).
In the end,
we successfully analyzed and generated reports
for 312 projects.
The build of the remaining projects failed
because it required human intervention,
e.g., to set up a specific environment
for the build.
This was also observed in prior work~\cite{java-builds}.

For diversity,
we discovered Make projects
from two sources.
First,
we used the Github {\sc api}
to collect popular C/C++ projects.
Second,
to ensure the buildability of the examined projects,
we also employed the Ultimate Debian Database
({\sc udd})~\cite{udd}
to identify widely-used Debian packages
based on the {\it ``vote''} metric,
which indicates the number of people
who regularly use a specific package~\cite{pop-con}.
The build workflow of Debian packages
uses the {\tt sbuild} utility~\cite{sbuild},
which automates the build process
of Debian binary packages
by creating the necessary build environment
(e.g., it installs all build dependencies
in an isolated environment)
for a particular architecture (e.g., x86-64).
{\tt sbuild} allows us to hook over
the build phase of its process.
In this manner we can monitor each build
and perform our own analysis.
We built every Debian package using
our Make wrapper
(Section~\ref{sec:implementation})
instead of the default Make command.
In total,
we examined 300 Make projects
coming from the Github and Debian ecosystems.
Overall,
the list of the selected Gradle and Make projects contains
popular ones
(e.g., the SQLite database,
the Spring framework,
and more)
which involve complex build scripts.
The characteristics of projects
are summarized in Table~\ref{table:eff}.

We ran every Gradle and Make build in sequential mode
as in the work of~\citet{nandor2019} to make fair
comparisons against mkcheck.
However,
our approach is able to support parallel builds by
tracking the thread (and its descendants) where
every build task is running.
Finally,
we ran the builds on Docker containers
inside a host machine with an Intel i7 3.6{\sc gh}z processor
with 8 cores and 16{\sc gb} of {\sc ram}.

\subsection{RQ1: Fault Detection Results}
\label{sec:eval-faults}

\begin{table}[t]
\centering
\caption{Faults detected by our approach.
Each table entry indicates
the number of faulty projects/the total
number of the examined projects (Projects),
the average LoC (Avg. LoC),
and the average lines
of build scripts (Avg. BLoC).
The columns MIN, MOUT,
and OV indicate
the number of projects where
missing inputs, missing outputs,
and ordering violations appear respectively.
}
\label{table:eff}
\resizebox{0.7\linewidth}{!}{%
\begin{tabular}{lrrr|rrr}
\hline
& \multicolumn{3}{c|}{{\bf Project Characteristics}}
& \multicolumn{3}{c}{\bf Fault types} \bigstrut \\
{\bf Build System} & {\bf Projects} & {\bf Avg. LoC}
& {\bf Avg. BLoC} & {\bf MIN}   & {\bf MOUT} & {\bf OV} \bigstrut\\
\hline
Gradle       & 73/312       & \nnum{35536}  & 589         & 58  & 20    & 25 \\
Make         & 251/300      & \nnum{74414} & \nnum{838}   & 249  & -    & 5 \\
\hline
\end{tabular}}
\vspace{-3mm}
\end{table}

Table~\ref{table:eff} summarizes
our fault detection results.
Our method identified problematic builds in
73 out of 312 Gradle projects.
There are 157 issues
related to incremental builds from which
122 faults are missing inputs
appearing in 58 projects,
while the remaining faults (35)
are associated with missing outputs
found in 20 projects.
Faulty parallel builds are also common in Gradle projects,
as we uncovered 80 ordering violations
in 25 Gradle repositories.
Furthermore,
our tool detected issues in 251 Make projects;
it discovered~\nnum{15740} Make target rules with missing inputs.
Most of them involved missing header dependencies
concerning object files.
It also reported~\nnum{14} ordering violations
that may lead to race conditions in 5 projects.
Note that missing outputs are only relevant to Gradle
(Section~\ref{sec:build-systems}).
For this reason,
we modeled every Make rule as a~\bfs~task
that was producing $\top$ (i.e., any file).
Therefore,
based on the definition of
the subsumption relation
(Figure~\ref{fig:subsume}),
a missing output issue (Definition~\ref{def:mout})
is not possible for Make.

To verify that the issues detected
by our tool are indeed faults,
we worked as follows.
For Gradle projects,
we examined each fault report,
and tried to reproduce it.
Specifically,
we automatically verified each issue
related to incremental builds by
checking that re-running Gradle does not
trigger the execution of tasks
marked with missing inputs/outputs by our tool,
even after updating the contents of
their dependent files.
We followed the same automated approach
for the verification of the reported Make faults
associated with incremental builds.
For ordering violations,
we manually verified that executing
conflicting build tasks in the erroneous order
can affect the outcome of a build,
e.g., causing build failures,
or producing build targets
with incorrect contents.

For Make projects,
we also analyzed every build with mkcheck,
which is the state-of-the-art tool
for Make-based builds~\cite{nandor2019}.
We then cross-checked
the results generated by our tool
with those produced by mkcheck
(see Section~\ref{sec:eval-comparison}).

\subsection{RQ2: Fault Importance}
\label{sec:eval-devs}

\begin{table}[t]
\caption{Faults confirmed and fixed by the developers.}
\label{table:fixed}
\resizebox{0.55\linewidth}{!}{%
\begin{threeparttable}[b]
\begin{tabular}{ll|rrrr}
\hline
{\bf Project}                   & {\bf Build System} & {\bf Total}
& {\bf MIN} & {\bf MOUT} & {\bf OV}\bigstrut \\
\hline
junit-reporter                  & Gradle        & 12    & 3   & 3    & 6  \\
muwire                          & Gradle        & 9     & 0   & 0    & 9  \\
aeron                           & Gradle        & 4     & 1   & 1    & 2  \\
conductor                       & Gradle        & 4     & 0   & 1    & 3  \\
xtext-gradle-plugin             & Gradle        & 4     & 0   & 0    & 4  \\
rundeck                         & Gradle        & 3     & 1   & 1    & 1  \\
apina                           & Gradle        & 3     & 0   & 0    & 3  \\
caffeine                        & Gradle        & 2     & 0   & 0    & 2  \\
RxAndroidBle                    & Gradle        & 2     & 0   & 0    & 2  \\
jmonkeyengine                   & Gradle        & 1     & 0   & 1    & 0  \\
tsar                            & Make          & 31    & 31  & -    & 0  \\
Cello                           & Make          & 27    & 27  & -    & 0  \\
webdis                          & Make          & 27    & 27  & -    & 0  \\
density                         & Make          & 17    & 17  & -    & 0  \\
VRP-Tabu                        & Make          & 12    & 12  & -    & 0  \\
pspg                            & Make          & 9     & 9   & -    & 0  \\
janet                           & Make          & 8     & 8   & -    & 0  \\
parcellite                      & Make          & 8     & 8   & -    & 0  \\
hdparam                         & Make          & 7     & 7   & -    & 1  \\
reptyr                          & Make          & 5     & 5   & -    & 0  \\
Others\tnote{1}                 & Make,Gradle   & 40    & 33  & 1    & 5  \\
\hline
                                & {\bf Total}   & {\bf 235} & {\bf 189}
								& {\bf 8} & {\bf 38} \bigstrut \\
\hline
\end{tabular}
\begin{tablenotes}
\item[1] \parbox[t]{\linewidth}{\small {\bf Others:} gps-sdr-sim, cscout, proxychains,
gradle-scripts, p2rank, goomph, groocss,
ShellExec, anna, fulibGradle, nf-tower, alfresco-gradle-sdk,
GradleRIO, helios, kscript, coveralls-gradle-plugin,
gradle-test-logger-plugin, joystick, cqmetrics,
JenkinsPipelineUnit, swagger-gradle-plugin,
gradle-swagger-generator-plugin,
gradle-sora, tinyrenderer, libcs50
}
\end{tablenotes}
\end{threeparttable}}
\vspace{-5mm}
\end{table}

We provided fixes for 71 Make and Gradle projects
that we chose while we were examining
their fault detection results,
and in turn,
we submitted patches to the upstream developers.
Patch generation was done manually,
mostly because of the complex structure of the faulty projects,
and the peculiar semantics of build systems' {\sc dsl}
(especially that of Gradle).
We leave repairing build scripts
through automated means as future work.

Table~\ref{table:fixed} enumerates the faults
that are confirmed and fixed.
Notably,
235 issues found in 45 out of 71 projects were fixed,
while most of the remaining patches are in a pending state.
The list of projects where
our patches were accepted contains
popular projects,
such as
{\tt tinyrenderer} (\textasciitilde8k stars),
{\tt caffeine} (> 7k stars),
{\tt aeron} (\textasciitilde5k stars),
{\tt Cello} (\textasciitilde5k stars),
and more.
The list also includes
projects that are maintained and developed
by well-established organisations,
such as {\tt conductor} (developed by Netflix),
{\tt tsar} (developed by the Alibaba Group).
This indicates that the faults we identified
do matter to the community.

\subsection{RQ3: Fault Patterns}
\label{sec:eval-patterns}

When we manually examined the issues
generated by our tool,
we recognized the following
five fault patterns,
which result in build failures,
time-consuming builds,
or erroneous build outcomes.
We identified three kinds of faults
related to incremental builds caused
by missing inputs or outputs.

{\bf Test resources.}
To ensure the correctness
of their programs,
developers typically specify dedicated build rules
for performing different forms of testing
(e.g., unit and functional testing)
during build.
Running tests is a time-consuming task~\cite{practical},
so the build rules associated with tests
are triggered only
when there are updates to any of
the source files that tests rely on.
As with source files,
changing any of the resources used by tests
(e.g., test data or additional helper scripts)
must re-run tests to make sure
that the change does not break anything.
Not running tests is a missed opportunity
to identify potential issues and may lead to
late identification of bugs.

For example,
the Gradle project {\tt kscript}
contains a test suite of Kotlin files
included in the {\tt test/resources} directory.
The tests of {\tt kscript} contains test assertions
that rely on the state and contents of the files included
in this test suite.
However,
the developers failed to declare the
test suite directory as input of
the Gradle task {\tt test}.
Our tool detected this fault,
and we reported to the developers
who fixed it.

{\bf Stale artifacts.}
As already discussed,
the main goal of a build is to construct artifacts,
such as executables,
libraries,
documentation accompanying software,
and more.
The build process must re-generate these artifacts,
when any of the files used for their construction is updated
since the last build.
Failing to do so can lead to stale artifacts,
which in turn,
can either harm the reliability of applications,
(e.g., cause runtime errors),
or generate wrong build outputs.

This pattern is particularly common in Make builds
where developers do not enumerate
the dependencies of object files correctly.
As an example of stale artifacts,
recall the build script of Figure~\ref{fig:motiv-incr}.
This example demonstrates
that even best practices for tracking dependencies automatically
(e.g., through {\tt gcc -MD}) are not sufficient
for ensuring the correctness of builds.

{\bf Time consuming tasks.}
The purpose of incremental builds is
to reduce build time
by running only the build tasks
needed to achieve a specific goal.
This boosts productivity
as it enables developers to get feedback
and respond to changes of their codebase much earlier.
To avoid unnecessary computation,
it is important
that time consuming build tasks are incremental.

We identified an instance of
this pattern in the Gradle project {\tt junit-reporter}.
This project contains some JavaScript source files
used to visualize JUnit reports in {\sc html} format.
Developers define a Gradle task
to bundle JavaScript source files
into a single file ({\tt site.js})
that is finally incorporated in
the {\sc html} page of test reports.
However,
this task was not declared as incremental.
As a result,
Gradle was bundling JavaScript files
at every build,
causing the subsequent Gradle tasks
that were dependent on {\tt site.js}
to be executed,
as Gradle was creating a newer version of {\tt site.js} each time.
Our tool marked {\tt site.js}
as missing output of the bundler task.
Based on this,
we sent a patch to the upstream developers
who integrated it in their codebase.
Fixing this issue made builds eight times faster.

\begin{figure}
\begin{lstlisting}[language=Make]
LIBS := $(addprefix build/lib/, $(LIB_BASE) $(LIB_VERSION))
$(LIBS): $(SRC)
  $(CC) $(CFLAGS) -o $(LIB_VERSION) $(SRC)
  $(CC) $(CFLAGS) -c -o $(LIB_OBJ) $(SRC)
  rm -f $(LIB_OBJ)
  ln -sf $(LIB_VERSION) $(LIB_BASE)
  mv $(LIB_VERSION) $(LIB_BASE) build/lib
\end{lstlisting}
\vspace{-3mm}
\caption{A Make script with conflicting producers.}
\label{fig:con-prod}
\vspace{-5mm}
\end{figure}

Below we discuss two categories of faults
related to parallel builds.

{\bf Conflicting Producers.}
We have identified issues
associated with tasks
that produce the same file
or write to the same output directory.
Parallel execution of such build tasks is harmful,
because race conditions may emerge,
when two tasks affecting the same state
(i.e., files) run concurrently.


The Gradle script of Figure~\ref{fig:motiv-paral}
is an example of conflicting producers.
Figure~\ref{fig:con-prod} shows
another example
coming from the {\tt libcs50} project.
This Make script defines a rule (line 2)
that creates two libraries
inside the {\tt build/lib} directory
(see variable {\tt \$(LIBS)}).
The code first compiles the source file
into the corresponding object file (line 3)
from which a shared library,
namely {\tt \$(LIB\_BASE)},
is constructed (line 4).
Then,
it creates a symbolic link ({\tt \$(LIB\_VERSION)})
pointing to the newly-created library (line 6),
and finally moves these files to
the {\tt /build/lib} directory (line 7).
The official documentation of {\sc gnu} Make states
that such a rule definition is incorrect~\cite{make-outs}.
In our example,
the rule at line 2 is executed twice
(one for every target defined in the {\tt \$(LIBS)})
variable).
Consequently,
the parallel build might crash
with the error
{\it ``mv: cannot stat 'libcs50.so.10.1.0': No such file or directory''},
as every rule execution races against each other.
Specifically,
when the second rule invocation attempts to
move the libraries,
they may have already been moved to {\tt /build/lib}
by the first rule.
The developers of {\tt libcs50}
immediately fixed this problem.

\begin{figure}
\begin{lstlisting}[language=gradle]
sourceSets { main { srcDir "build/generated-sources/" } }
task generateNodes(type: JavaExec) {
  main = "com.github.benmanes.caffeine.cache.NodeFactoryGenerator"
  args "build/generated-sources/"
  outputs.dir "build/generated-sources/" }
apply plugin: "com.bmuschko.nexus"
\end{lstlisting}
\vspace{-3.5mm}
\caption{A Gradle script manifesting an ordering violation.}
\label{fig:gen-files}
\vspace{-5.5mm}
\end{figure}

{\bf Generated Source Files and Resources.}
Many projects generate part of their source code
or resources at build time.
These automatically generated source files
and resources are then compiled
or used later by other build tasks
to form the final artifacts of the build process,
e.g., binaries.
Developers must be careful enough
to preserve the correct execution order
between the build tasks
that are responsible for generating
and using these source files and resources.
Ordering violations
(e.g., compiling code when source files are missing)
are the root cause for build failures,
or subtle errors detected
at a later stage of software lifecycle.

Figure~\ref{fig:gen-files} presents
a code fragment taken
from the popular {\tt caffeine} project.
The code specifies
that the source files of the project
are stored in the {\tt build/generated-sources} directory
(line 1).
These source files are generated automatically
by the Gradle task {\tt generateNodes}.
To do so,
this task runs the class {\tt NodeFactoryGenerator}
with {\tt "build/generated-sources"} as an argument
(lines 2--5).
Then,
this code applies the plugin {\tt "com.bmuschko.nexus"}
used for uploading the sources {\sc jar} file
to a remote repository.
To assemble a {\sc jar} file
containing the source files of the application,
this plugin adds the {\tt sourcesJar} task
to the project.
The problem with this code is that
no dependency is declared between the tasks
{\tt generateNodes} and {\tt sourcesJar}.
Thus,
the build process uploads empty artifacts
to the remote repository,
when Gradle executes {\tt sourcesJar} before {\tt generateNodes}.
The developers of {\tt caffeine}
confirmed and fixed this ordering issue.

\subsection{RQ4: Performance}
\label{sec:eval-performance}

To measure the performance of our approach
we recorded the time spent at each step
(recall Figure~\ref{fig:arch}).
The generation step,
which is responsible for executing
and monitoring our instrumented builds,
dominates the execution time of our method.
In particular,
this step slows down
both Gradle and Make builds
by a factor of around two for the 90$^\text{th}$ percentile
of the examined projects.
This is consistent with
the recent literature~\cite{nandor2019},
as the main overhead of
this phase stems from the system call tracing utility
(i.e., {\tt strace}).

The analysis of~\bfs~programs takes around 2.47
and 5.1 seconds
on average for Make and Gradle projects respectively,
and is linear to the size of programs.
This phase is efficient enough
to analyze {\sc gb}s of programs in
a reasonable time (e.g., 6.9{\sc gb} in less
than 3 minutes).
In online mode,
though,
the observable time spent on
the analysis step is eliminated,
as the overall time is bounded to the time needed
for a build.
As explained in~\ref{sec:implementation},
this is because the processing of~\bfs~programs is faster
than the build itself,
and thus we take advantage of multicore architectures.
Finally,
the fault detection step is pretty fast;
it takes only 0.11 and 0.45 seconds
on average for Gradle and Make projects respectively.

\subsection{RQ5: Comparison with state-of-the-art}
\label{sec:eval-comparison}

\begin{figure}[t]
\hspace{1mm}
\begin{minipage}[b]{0.45\linewidth}
\captionof{table}{Time spent on analyzing builds
and detecting faults
by our approach vs. mkcheck (in seconds).}
\vspace{3mm}
\label{table:times}
\resizebox{\linewidth}{!}{%
\begin{tabular}[b]{l|rr|rr}
\hline
& \multicolumn{2}{c|}{{\bf Median}}
& \multicolumn{2}{c}{\bf Average} \bigstrut \\
{\bf Phase} & {\bf BuildFS} & {\bf mkcheck}
& {\bf BuildFS} & {\bf mkcheck} \bigstrut\\
\hline
Build              & 4.1  & 4.9    & 20.5 & 22.2   \\
Fault detection    & 0.01 & 182.62 & 0.45  & \nnum{3368} \\
Overall            & 4.1  & 186.75 & 21   & \nnum{3390} \\
\hline
\end{tabular}}
\end{minipage}
\hfill
\hspace{1.5mm}
\begin{minipage}[b]{0.51\linewidth}
\includegraphics[scale=0.35]{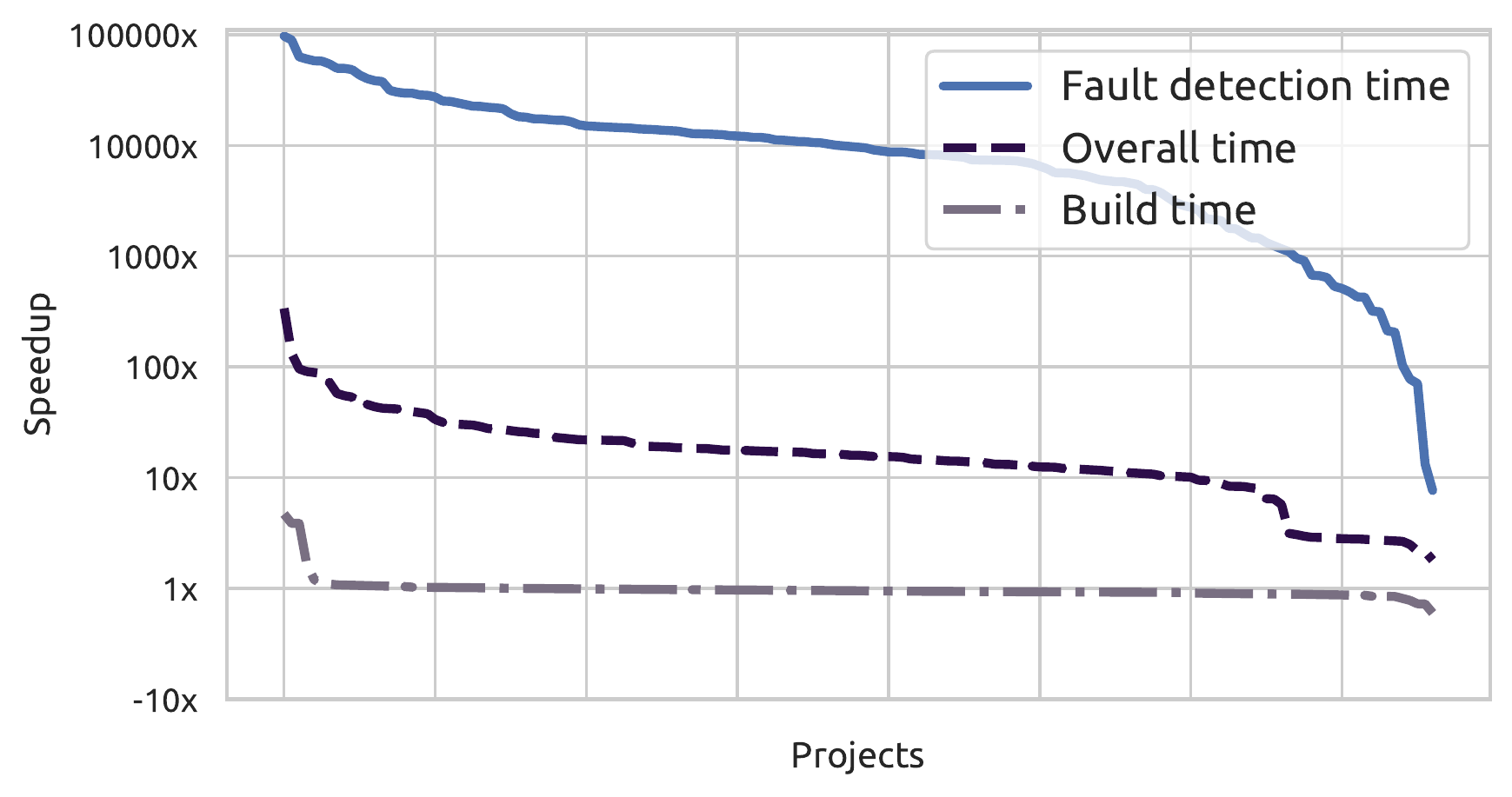}
\captionof{figure}{The speedup for every Make project
by\\ our approach over mkcheck.}
\label{fig:speedup}
\end{minipage}
\end{figure}

As a first step,
we built and analyzed a number of
Gradle projects with mkcheck.
After finding that mkcheck produces
meaningless reports
and an overwhelming number of false positives
(this is not surprising because mkcheck is unable
to deal with Java-based builds
as explained in Section~\ref{sec:motivation}),
we focused on performing comparisons only
for Make projects.
We applied mkcheck to the 300 Make projects
and we recorded the fault reports
and the time spent at each phase
(i.e., build time and fault detection time).
Note that build time includes the time needed
for building project
as well as
the time taken for generating
and analyzing the build trace.

In terms of fault detection,
mkcheck produced false positives in three cases
due to granularity of processes.
That is, two build tasks were merged into a single task
because they were governed
by the same system process,
leading to imprecision.
False positives are also observed
in the initial work of~\citet{nandor2019}.
On the contrary,
our approach did not generate false positives
as it can reliably determine all file accesses
of each task as explained
in Section~\ref{sec:gen-traces}.

Figure~\ref{fig:speedup} demonstrates
the relative times between
building and detecting faults with our approach
and mkcheck.
Notice
that our tool and mkcheck spend
almost the same amount of time
for building and monitoring a project.
The maximum speedup is 9x,
the minimum is -1.25x,
while the average is 1.19x.
Although mkcheck uses {\tt ptrace}
for tracking system operations,
which is 20\% times faster than {\tt strace}~\cite{nandor2019},
our approach benefits
from running the generation
and analysis steps concurrently.
Moving to fault detection times,
we observe that our approach is much faster.
Specifically,
we can detect faulty builds
up to six orders of magnitude
faster than mkcheck,
and the minimum speedup is only 83x.
This huge speedup is explained by the fact
that our approach needs~\emph{only}
one build to uncover faults.
Notably,
in projects consisting of a large number of source files,
mkcheck required days to detect faults
(e.g., it spent 3.3 days for
analyzing {\tt ghostscript}).
This is because mkcheck performs an incremental build
per source file to verify correctness,
something that hinders its scalability.
Finally,
when considering the overall time
(i.e., build + fault detection time),
our approach is 74x times faster than mkcheck,
on average,
while the maximum and minimum speedup is
\nnum{1837}x and 2.6x respectively.

We also provide some absolute performance times on
Table~\ref{table:times}.
Overall,
our method required~21 seconds (on average) for analyzing
builds and detecting faults
(see phase ``Overall'' on Table~\ref{table:times}),
while mkcheck spent~\nnum{3390} seconds on average
for performing the same tasks.
The median time is~4.1 and~186.75 seconds for our tool
and mkcheck respectively.
Notably,
mkcheck spent more than ten minutes for reporting faults
in the 29\% of the inspected Make projects.

These findings indicate that our method is superior to
the state-of-the-art in terms of both fault detection
and performance.
Moreover,
due to its effectiveness and efficiency,
we argue that our method can be used in practice
as part of the software testing pipeline.

\section{Related Work}
\label{sec:related}

Our work is related to
four research areas:
testing and debugging builds,
understanding and refactoring builds,
trace analysis,
and regression test selection.

{\bf Testing and Debugging Builds.}
Testing build scripts is an emerging research area.
mkcheck~\cite{nandor2019},
and {\sc bee}~\cite{bee} are two tools
that also detect missing inputs,
but they are tailored for Make-based builds.
As we pointed out in~\ref{sec:motivation},
these tools have two important limitations
that concern:
(1) low precision
when applied to Java build tools,
and (2) efficiency \& applicability.
As we explained earlier,
our approach tackles both limitations.

Beyond testing,
a number of studies have been developed
to identify the root causes of problematic builds,
and suggest fixes for them.
\citet{build-fault},
have designed a tool that
given a failed build,
it identifies the faulty Make rules
that caused the build crash.
Their approach performs
an instrumentation on a Make build
that tracks the execution trace of each Make rule,
and records the crash point of the build.
Based on a probabilistic model,
they assign different scores to every rule,
indicating the probability
that the rule caused the crash.
Subsequent work~\cite{unreproducible,unreproducible2}
have focused on locating faults of unreproducible builds.
Reproducibility is a property
that ensures that
a build is deterministic
and always results in bitwise-identical targets
given the same sources
and build environment.
The initial work of~\citet{unreproducible}
analyzes the logs from an unreproducible build,
and proposes a ranked list
of problematic source files
that might contain the fault.
Recently,
the authors extended their approach~\cite{unreproducible2}
to locate the specific command
that is responsible for
the unreproducible build.
To do so,
they employed a backtracking analysis
on system call trace
stemming from build execution.
Contrary to these approaches,
our method does not require a build failure,
but it is capable of detecting latent future faults.

{\bf Understanding and Refactoring Builds}.
There are plenty of tools
developed over the past decade
to assist developers in understanding
and refactoring builds.
{\it Makao}~\cite{makao} is a Make-related framework
used for visualizing build dependencies.
By extracting knowledge
from such dependencies
through filtering and querying,
Makao provides support for
refactoring build scripts
via an aspect-oriented approach.
{\it SYMake}~\cite{symake},
evaluates Makefiles
and produces
(1) a symbolic dependency graph,
and (2) a symbolic execution trace.
Then,
it applies different algorithms
to the results to detect a
number of code smells
(e.g., cyclic dependencies),
and perform refactoring on Make scripts
(e.g., target renaming).
{\sc Metamorphosis}~\cite{metamorphosis}
is a tool used to migrate existing build scripts
to {\sc CloudMake}~\cite{cloudmake},
which is a modern build system
developed by Microsoft.
As a starting point,
{\sc Metamorphosis} analyzes
the execution trace of a given build
and then automatically synthesizes
an initial {\sc CloudMake} script
that reflects
the behavior of the original script.
Then,
it optimizes the build script
synthesized by the previous step
by applying a sequence of transformations
and choosing the best possible ones
based on a fitness function.
\citet{decomposition}
propose a new refactoring method,
target decomposition,
for dealing with underutilized targets;
a build-related code smell
that causes slower builds,
larger binaries,
and less modular code.

{\bf Trace Analysis.}
Most of the existing
work~\cite{metamorphosis,license,nandor2019,grex,unreproducible2,puppet-paper}
that is relevant to the domain of builds
employs techniques for analyzing traces---and
especially system call traces.
Our work differs from the previous approaches
as the proposed model (\bfs),
and in turn,
its practical realization
captures both the dynamic behavior
and the static specification
of high-level programming constructs
(i.e., build tasks).
This enables us to verify---contrary to
existing approaches---the execution
of a build phase with regards to
its specification,
while monitoring build,
making our method more precise,
efficient,
and generally-applicable.

{\bf Regression Test Selection.}
Recently,
there have been advances on
dynamic regression test selection techniques
({\sc rts})~\cite{practical,refactoring-rts,jvm,hybrid}.
Dynamic {\sc rts} methods improve
the performance of regression testing
by running only those tests
affected by a specific code change.
To do so,
they compute test dependencies
from previous test runs.
\citet{practical},
and~\citet{jvm} extract
test dependencies by determining
the execution boundaries of each test,
and tracking all intermediate file accesses.
{\sc rts} methods and our technique are complementary;
they can both used as part of a build
to improve efficiency and reliability respectively.

\section{Conclusion}
\label{sec:conclusions}

We developed
a generic and practical approach
for discovering faults that can
cause incremental and parallel build failures.
To do so,
we proposed a model (\bfs)
for arbitrary build executions
that captures
the static specification
and the dynamic behavior
of each build task.
We then formally defined three types
of faults concerning
incrementality and parallelism,
and presented an approach
for exploiting~\bfs~and
detecting such faults in practice.
Combining static and dynamic information
in a single representation made
our method efficient
and applicable to any build system.

Our method was able to uncover issues
in hundreds of Make and Gradle builds.
Notably,
our approach tackled the limitations of existing work,
and it is the first to deal with Java-based build tools.
We demonstrated the importance of the discovered faults
by providing patches to numerous projects.
Thanks to our tool,
the developers of 45 open-source projects
confirmed and fixed 235 issues,
in total.
Moreover,
a comparison between our tool
and a state-of-the-art Make-based tool showed
that our approach is more effective
and orders of magnitude faster.
We argue that
our tool could be part of the
software testing pipeline,
helping developers to discover defects
and inconsistencies in
their software artifacts
that arise from faulty build definitions.

\bibliography{main}

\end{document}

%% file: includes.tex
\usepackage{url}
\usepackage{graphicx}
\usepackage{listings}
\usepackage{xcolor}
\usepackage{subcaption}
\usepackage{enumitem}
\usepackage{amsmath}
\usepackage{bigstrut}
\usepackage{mathpartir}
\usepackage[altpo]{backnaur}
\usepackage{color}
\usepackage{cleveref}
\usepackage{adjustbox}
\usepackage{nccmath}
\usepackage{bbding}
\usepackage{sepnum}
\usepackage{caption}
\usepackage{subcaption}
\usepackage{threeparttable}
\usepackage[linesnumbered,ruled]{algorithm2e}
\usepackage{natbib}

\setcitestyle{authoryear,citesep={;}}

\SetCommentSty{mycommfont}

\PassOptionsToPackage{hyphens}{url}\usepackage{hyperref}

\setlist[itemize]{noitemsep, leftmargin=1em}
\setlist[enumerate]{noitemsep, leftmargin=1em}
\newcommand{\nnum}[1]{\sepnum{.}{,}{}{#1}}

\makeatletter
\def\@copyrightspace{\relax}
\makeatother

\newcommand{\bfs}{{\sc b}uild{\sc fs}}

\newcommand{\prim}[1]{\text{{\sffamily{\small {\tt #1}}}}}
\newcommand{\primsem}[1]{\text{{\sffamily{\footnotesize {\tt #1}}}}}
\newcommand{\primarr}[1]{\text{{\sffamily{\tiny {\tt #1}}}}}

\setlist[itemize]{noitemsep, leftmargin=1em}
\setlist[enumerate]{noitemsep, leftmargin=1em}

\definecolor{dkred}{RGB}{87,10,10}
\definecolor{burntorange}{rgb}{0.8, 0.33, 0.0}
\definecolor{OrangeRed}{rgb}{1.0, 0.27, 0.0}
\definecolor{ForestGreen}{rgb}{0.0, 0.27, 0.13}

\lstdefinelanguage{make}{
  morecomment = [l]{;},
  morecomment=[s]{/*}{*/},
  morestring=[b]",
  sensitive = true,
  classoffset=0,
  morecomment=[l]\#,
  morekeywords={
      -,
      include,
  },
  classoffset=1, keywordstyle=\ttfamily\color[rgb]{0,0,1},
  morekeywords={
      \$@,
      \$,
      \textasciicircum
  },
  alsoletter={\%},
  keywordsprefix={\%},
}

\lstdefinelanguage{gradle}{
  morecomment = [l]{;},
  morecomment=[s]{/*}{*/},
  morestring=[b]",
  sensitive = true,
  classoffset=0,
  morecomment=[l]//,
  morekeywords={
      task,
      doFirst,
      shadowJar,
      from,
      into,
      new,
      def,
      plugin,
      :
  },
  classoffset=1, keywordstyle=\ttfamily\color[rgb]{0,0,1},
  morekeywords={
  },
  alsoletter={\%},
  keywordsprefix={\%},
}

\lstdefinelanguage{fstrace}{
  morestring=[b]",
  sensitive = true,
  classoffset=0,
  classoffset=1,
  morecomment=[l]\#,
  keywordstyle=\bfseries\sffamily\footnotesize\color[rgb]{0.0,0.0,0.0},
  stringstyle=\ttfamily\color[rgb]{0,0,1},
  morekeywords={
  },
  alsoletter={\%},
  keywordsprefix={\%},
}

\lstdefinelanguage{strace}{
  emph={echo},
  emphstyle={\color{OrangeRed}},
  morestring=[b]",
  sensitive = true,
  classoffset=0,
  classoffset=1,
  morecomment=[l]+,
  keywordstyle=\sffamily\color[rgb]{0.627,0.126,0.941},
  stringstyle=\ttfamily\color[rgb]{0,0,1},
  morekeywords={
  },
  alsoletter={\%},
  keywordsprefix={\%},
}

\lstdefinelanguage{Kotlin}{
  comment=[l]{//},
  emph={delegate, filter, first, firstOrNull,  lazy, map, mapNotNull, println, return@},
  emphstyle={\color{OrangeRed}},
  identifierstyle=\color{black},
  keywords={abstract, actual, as, as?, break, by, class, companion, continue, data, do, dynamic, else, enum, expect, false, final, for, fun, get, if, import, in, interface, internal, is, null, object, override, package, private, public, return, set, super, suspend, this, throw, true, try, typealias, val, var, vararg, when, where, while},
  keywordstyle={\color{NavyBlue}\bfseries},
  morecomment=[s]{/*}{*/},
  morestring=[b]",
  morestring=[s]{"""*}{*"""},
  ndkeywords={@Deprecated, @JvmField, @JvmName, @JvmOverloads, @JvmStatic, @JvmSynthetic, Array, Byte, Double, Float, Int, Integer, Iterable, Long, Runnable, Short, String},
  ndkeywordstyle={\color[rgb]{0,0,0}\bfseries},
  sensitive=true,
  stringstyle={\color[rgb]{0,0,1}\ttfamily},
}

%% file: syntax.tex
\begin{figure}[t]
\centering
\begin{subfigure}{.4\linewidth}
\footnotesize
\begin{bnf*}
    \bnfprod{$b \in {\it Build}$}
    {\bnftd{$t^{*}$}}
    \hspace{13.9mm}
    \bnfts{[build execution]}\\[-0.5mm]
    \bnfprod{$t \in {\it Task}$}
    {\bnftd{\primsem{task} $\tau$ $k$: $k$ \primsem{after} $d$ = $s$}}
    \hspace{4mm}
    \bnfts{[task]}\\[-0.5mm]
    \bnfprod{$d \in {\it Dep}$}
    {\bnftd{$\bot$} \bnfor \bnftd{$\tau$} \bnfor \bnftd{($\tau\dots$)}}
    \hspace{7.6mm}
    \bnfts{[task deps]}\\[-0.5mm]
    \bnfprod{$k \in {\it FileSpec}$}
    {\bnftd{$p$}
    \bnfor \bnftd{($p\dots$)}
    \bnfor \bnftd{$\bot$}
    \bnfor \bnftd{$\top$}}
    \hspace{3.2mm}
    \bnfts{[file spec]}\\[-0.5mm]
    \bnfprod{$s \in {\it Statements}$}
    {\bnftd{\primsem{sysOp in} $z$: $o$}}
    \hspace{8.3mm}
    \bnfts{[operation]}\\[-0.5mm]
    \bnfmore{\bnfor \bnftd{\primsem{newproc} $z$}}
    \hspace{8mm}
    \bnfts{[new process]}\\[-0.5mm]
    \bnfmore{
     \bnfor \bnftd{\primsem{newproc} $z$ \primsem{from} $z$}
    }
    \hspace{9.2mm}
    \bnfts{[fork]}\\[-0.5mm]
    \bnfmore{\bnfor \bnftd{$s$;$s$}}
    \hspace{14.1mm}
    \bnfts{[compound stmt]}\\[-0.5mm]
    \bnfprod{$o \in {\it Op}$}
    {\bnftd{$\primsem{let fd}_f$ = $e$}}
    \hspace{10.3mm}
    \bnfts{[create fd]}\\[-0.5mm]
    \bnfmore{\bnfor \bnftd{\primsem{del}($\primsem{fd}_f$)}}
    \hspace{11.2mm}
    \bnfts{[destroy fd]}\\[-0.5mm]
    \bnfmore{\bnfor \bnftd{\primsem{consume}($e$})}
    \hspace{5.6mm}
    \bnfts{[consume path]}\\[-0.5mm]
    \bnfmore{\bnfor \bnftd{\primsem{produce}($e$)}}
    \hspace{5.7mm}
    \bnfts{[produce path]}\\[-0.5mm]
    \bnfmore{\bnfor \bnftd{$o$;$o$}}
    \hspace{16.55mm}
    \bnfts{[compound op]}\\[-0.5mm]
    \bnfprod{$e \in {\it Expr}$}
    {\bnftd{$p$}}
    \hspace{30.2mm}
    \bnfts{[path]}\\[-0.5mm]
    \bnfmore{\bnfor \bnftd{$\primsem{fd}_f$}}
    \hspace{28.6mm}
    \bnfts{[fd]}\\[-0.5mm]
    \bnfmore{\bnfor \bnftd{$p$ $\primsem{at } e$}}
    \hspace{6.3mm}
    \bnfts{[$p$ relative to $e$]}\\[-0.5mm]
    \bnfprod{$p \in {\it Path}$}
    \bnftd{\text{is the set of paths}}\\[-0.5mm]
    \bnfprod{$\tau \in {\it TaskName}$}
    \bnftd{\text{is the set of task names}}\\[-0.5mm]
    \bnfprod{$f \in {\it FileDesc}$}
    \bnftd{\text{is the set of file descriptors}}\\[-0.5mm]
    \bnfprod{$z \in {\it Proc}$}
    \bnftd{\text{is the set of processes}}\\[-0.5mm]
\end{bnf*}
\end{subfigure}
\begin{subfigure}{.5\linewidth}
\centering
\footnotesize
\begin{align*}
    \llbracket e \rrbracket &\in \pi \rightarrow \textit{Path} \\
    \llbracket p \rrbracket_{\pi} &\Rightarrow p \\
    \llbracket \primsem{fd}_f \rrbracket_{\pi} &\Rightarrow \pi(\primsem{fd}_f)\\
    \llbracket p\ \primsem{at}\ \primsem{fd}_f \rrbracket_{\pi} &\Rightarrow \textsc{join}(\pi(f), p) \\
    \llbracket o \rrbracket &\in  \pi \times r \rightarrow \pi \times r \\
    \llbracket \primsem{let fd}_f = e\rrbracket_{\pi,r} &\Rightarrow
    (\pi[f \rightarrow \llbracket e \rrbracket_\pi], r) \\
    \llbracket \primsem{del}(\primsem{fd}_f) \rrbracket_{\pi, r} &\Rightarrow
    (\pi[f \rightarrow \bot], r) \\
    \llbracket \primsem{consume}(e) \rrbracket_{\pi, r} &\Rightarrow
    (\pi, r_{\downarrow{\textit{c}}} \cdot \llbracket e \rrbracket_\pi) \\
    \llbracket \primsem{produce}(e) \rrbracket_{\pi, r} &\Rightarrow
    (\pi, r_{\downarrow{\textit{p}}} \cdot \llbracket e \rrbracket_\pi) \\
    \llbracket o_1;o_2\rrbracket_{\pi, r} &\Rightarrow
    \llbracket o_2 \rrbracket_{\pi', r'}\:\:\: (\pi', r') = \llbracket o_1 \rrbracket_{\pi, r} \\
    \llbracket s \rrbracket &\in \sigma \times r \rightarrow \sigma \times r \\
    \llbracket \primsem{sysOp in}\ p = s \rrbracket_{\sigma, r} &\Rightarrow
    (\sigma[p \rightarrow \pi], r)\:\:\: (\pi, r) = \llbracket s \rrbracket_{\sigma(p), r} \\
    \llbracket \primsem{newproc}\ p\rrbracket_{\sigma, r} &\Rightarrow
    (\sigma[p \rightarrow \bot], r) \\
    \llbracket \primsem{newproc}\ p_1\ \primsem{from} p_2\rrbracket_{\sigma, r} &\Rightarrow
    (\sigma[p_1 \rightarrow \sigma(p_2)], r) \\
    \llbracket s_1;s_2\rrbracket_{\pi, r} &\Rightarrow
    \llbracket s_2 \rrbracket_{\pi', r'}\:\:\: (\sigma', r') = \llbracket s_1 \rrbracket_{\sigma, r} \\
    \llbracket t \rrbracket &\in \sigma \rightarrow \sigma \times r \\
    \llbracket \primsem{task}\ \tau\ (k_1): k_2\ \primsem{after}\ d = s\rrbracket_\sigma &\Rightarrow
    \llbracket s \rrbracket_{\sigma, \bot}
\end{align*}
\end{subfigure}
\vspace{-3mm}
\caption{The syntax for representing build executions, along with
the semantics of~\bfs~expressions, operations, statements, and tasks.}
\label{fig:syntax}
\vspace{-6mm}
\end{figure}